\documentclass[a4paper, 12pt, openany, oneside]{article}
\usepackage[cp1251]{inputenc} 
\usepackage[english]{babel} 
\usepackage{amsmath, latexsym,  amssymb, array, graphics,
amsfonts, bm, eucal}

\usepackage[dvips]{graphicx}
\usepackage[flushleft,small]{caption2} 

\usepackage{indentfirst}

\makeatletter
\newcommand{\intl}{\int\limits}
\newcommand{\Res}{\mathop{\rm Res\,}}

\renewcommand{\Re}{\mathop{\rm Re\,}}
\newcommand{\ch}{\mathop{\rm ch}}
\newcommand{\sh}{\mathop{\rm sh}}
\renewcommand{\th}{\mathop{\rm th}}
\renewcommand{\Im}{\mathop{\rm Im\,}}
\newcommand{\sign}{\mathop{\rm sign\,}}
\makeatother
\usepackage{indentfirst}

\newcommand{\mc}[1]{\mathcal{#1}}
\newcommand{\E}{\mc{E}}

\makeatother

\usepackage[dvips]{graphicx}
\usepackage[flushleft,small]{caption2}

\begin{document}

\thispagestyle{empty}
\renewcommand{\abstractname}{\,Abstract}
\begin{flushright}\it\Large
Dedicated to the Memory of\\Anatoly Alexandrovich 
Vlasov\footnote{A.A. Vlasov worked managing chair of theoretical 
physics in Moscow State Regional
University in 50-60th years of last century}
\end{flushright}

 \begin{center}
\bf Oscillations of Degenerate Plasma in Layer with Specular --
Accommodative Boundary Conditions
\end{center}

\begin{center}
  \bf  A. V. Latyshev\footnote{$avlatyshev@mail.ru$} and
  A. A. Yushkanov\footnote{$yushkanov@inbox.ru$}
\end{center}\medskip

\begin{center}
{\it Faculty of Physics and Mathematics,\\ Moscow State Regional
University,  105005,\\ Moscow, Radio str., 10--A}
\end{center}\medskip

\tableofcontents
\setcounter{secnumdepth}{4}

\begin{abstract}
In the present paper the linearized problem of plasma oscillations
in layer (particularly, in thin films) in external longitudinal 
alternating electric field is solved
analytically. Specular -- accommodative  boundary conditions of
electron reflection from the plasma boundary are considered.
Coefficients of continuous and discrete spectra of the problem are
found, and electron distribution function on the plasma boundary and
electric field are expressed in explicit form. Absorption of energy 
of electric field in layer is calculated.

Refs. 34. Figs. 2.\\

\end{abstract}


\begin{center}
\item{}\section{Introduction}
\end{center}

\renewcommand{\refname}{\begin{center} \rm REFERENCES  \end{center}}

The present paper is devoted to degenerate electron plasma behaviour
research. Analysis of processes taking place in plasma under effect
of external electric field, plasma waves oscillations with various
types of condi\-tions of electron reflection from the boundary has
important significance today in connection with problems of such
intensively developing fields as microelec\-tronics and nanotechnologies
\cite{1} -- \cite{Liboff}.

The concept of "plasma"\, appeared in the works of Tonks and Langmuir
for the first time (see \cite{Langmuir}--\cite{Tonks2}),
the concept of "plasma frequency"\, was introduced in the same works
and first questions of plasma oscillations were considered there.
However, in these works equation for the electric field was
considered separately from the kinetic equation.

A.A. Vlasov \cite{Vlasov} for the first time introduced the concept
of "self-consistent electric field"\, and added the corresponding item
to the kinetic equation. Now equations describing plasma behaviour
consist of anchor system of equations of Maxwell and Boltzmann.
The problem of electron plasma oscillations was considered by A.A.
Vlasov \cite{Vlasov} by means of solution of the kinetic equation
which included self-consistent electric field.

L.D. Landau \cite{Landau} had supposed that outside of the half-space
containing degenerate plasma external electromagnetic field causing
oscillations in plasma is situated. By this Landau has formulated a
boundary condition on the plasma boundary. After that the problem
of plasma oscillation turns out to be formulated correctly as a
boundary value problem of mathematical physics.

In \cite{Landau}  L.D. Landau has solved analytically by Fourier
series the problem
of collisionless plasma behaviour in a half-space, situated in
external lon\-gi\-tu\-dinal (perpen\-di\-cular to the surface) electric
field, in conditions of specular reflection of electrons
from the boundary.

Further the problem of electron plasma oscillations was considered
by many authors. Full analytical solution of the problem is given
in the works \cite{8} and \cite{6}.

This problem has important significance in the theory of plasma (see,
for instance, \cite{2}, \cite{3} and the references in these works,
and also \cite{M1}, \cite{M2}).

The problem of plasma oscillations with diffuse boundary condition
was considered in the works \cite{Keller}, \cite{Kliewer} by method 
of integral transformations.
In the works \cite{Gohfeld1}, \cite{Gohfeld2} general asymptotic
analysis of electric field behaviour at the large distance from
the surface was carried out. In the work \cite{Gohfeld1} particular
significance of plasma behaviour analysis close to plasma re\-so\-nan\-ce
was shown. And in the same work \cite{Gohfeld1}  it was stated that
plasma behaviour in this case for conditions of specular and diffuse
electron scat\-te\-ring on the surface differs substantially.

In the works \cite{5} and \cite{5} general questions of this problem
solvability were considered, but diffuse boundary conditions were
taken into account. In the work \cite{5} structure of discrete
spectrum in dependence of parameters of the problem was analyzed.
The detailed analysis of the solution in general case in the works
mentioned above hasn't been carried out conside\-ring the complex
character of this solution.

The present work is a continuation of electron plasma behaviour in
external longitudinal alternating electric field research
\cite{5} -- \cite{11}.

In the present paper the linearized 
problem of plasma oscillations
in external alternating electric field in layer (particularly, in
thin films) is solved analytically.
Specular -- accommo\-da\-ti\-ve boundary conditions for electron
reflection from the boundary are considered. In \cite{9}--\cite{11}
diffuse boundary conditions were considered.

The coefficients of continuous and discrete spectra of the problem
are obtained in the present work, which allows us to derive
expressions for electron distribution function at the boundary
of conductive medium and electric field in explicit form, to reveal
the dependence of this expressions on normal momentum accommodation
coefficient and to show that in the case when normal electron
momentum accommodation coefficient equals to zero electron
distribution function and electric field are expressed by
known formulas obtained earlier in \cite{8}, \cite{6}.

The present work is a continuation of our work \cite{Gritsienko},
in which questions of plasma waves specular reflection 
from the plane boundary
bounding degenerate plasma were considered.

Let us note, that questions of plasma oscillations are also
considered in nonlinear statement (see, for instance, the work
\cite{Stenflo}, \cite{Stenflo2}).

\begin{center}
\item{}\section{Formulation of problem}
\end{center}

Let degenerate plasma occupy a slab  (particularly, thin films) $-a<x<a$.

We take system of equations describing plasma behaviour.
As a  kinetic equation we take Boltzmann --- Vlasov $\tau$--model 
kinetic equation:
$$
\dfrac{\partial f}{\partial t}+\mathbf{v}\dfrac{\partial f}{\partial
\mathbf{r}}+e\mathbf{E}\dfrac{\partial f}{\partial \mathbf{p}}=
\dfrac{f_{eq}(\mathbf{r},t)-f(\mathbf{r},\mathbf{v},t)}{\tau}.
\eqno{(1.1)}
$$

Here $f=f(\mathbf{r},\mathbf{v},t)$ is the electron distribution
function, $e$ is the electron charge, $\mathbf{p}=m\mathbf{v}$
is the momentum of an electron, $m$ is the electron mass, $\tau$
is the character time between two collisions,
$\mathbf{E}=\mathbf{E}(\mathbf{r},t)$ is the self-consistent
electric field inside plasma,
$f_{eq}=f_{eq}(\mathbf{r},t)$ is the local equilibrium Fermi --- Dirac distribution function,
$ f_{eq}=\Theta(\E_F(t,x)-\E), $ where $\Theta(x)$ is the function
of Heaviside,
$$
\Theta(x)=\left\{\begin{array}{c}
                   1, \quad x>0, \\
                   0,\quad x<0,
                 \end{array}\right.
$$
$\E_F(t,x)=\frac{1}{2}mv_F^2(t,x)$ is the disturbed kinetic energy
of Fermi, $\E=\frac{1}{2}mv^2$ is the kinetic energy of electron.

Let us take the Maxwell equation for electric field
$$
{\rm div}\,{\mathbf{E}(\mathbf{r},t)}= 4\pi\rho(\mathbf{r},t).
\eqno{(1.2)}
$$
Here $\rho(\mathbf{r},t)$ is the charge density,
$$
\rho(\mathbf{r},t)=e\int (f(\mathbf{r},
\mathbf{v},t)-f_{0}(\mathbf{v})) \,d\Omega_F, \eqno{(1.3)}
$$
where
$$
d\Omega_F=\dfrac{2d^3p}{(2\pi\hbar)^3}, \qquad
d^3p=dp_xdp_ydp_z.
$$

Here $f_{0}$  is the undisturbed Fermi --- Dirac electron 
distribution function,
$$
f_{0}(\E)=\Theta(\E_F-\E),
$$
$\hbar$ is the Planck's constant, $\nu$ is the effective frequency
of electron collisions, $\nu=1/\tau$, $\E_F=\frac{1}{2}mv_F^2$ is
the undisturbed kinetic energy of Fermi, $v_F$ is the electron
velocity at the Fermi surface, which is supposed to be spherical.

We assume that external electric field outside the plasma is
perpendi\-cu\-lar to the plasma boundary and changes according to
the following law: $E_{0}\exp(-i\omega t).$

Then one can consider that self-consistent
electric field
$\mathbf{E}(\mathbf{r},t)$ inside plasma has one
$x$--component and changes only lengthwise the axis $x$:
$$
\mathbf{E}=\{E_x(x,t),0,0\}.
$$

Under this configuration the electric field is perpendicular
to the boun\-da\-ry of plasma, which is situated in the plane $x=0$.

We will linearize the local equilibrium Fermi --- Dirac 
distribution $f_{eq}$
in regard to the undisturbed distribution $f_0(\E)$:
$$
f_{eq}=f_{0}(\E)+[\E_{F}(x,t)-\E]\delta(\E_{F}-\E),
$$
where $\delta(x)$ is the delta -- function of Dirac.

We also linearize the electron distribution function $f$ in terms of
absolute Fermi --- Dirac distribution
$f_{0}(\E)$:
$$
f=f_0(\E)+f_1(x,\mathbf{v},t).
\eqno{(1.4)}
$$

After the linearization of the equations (1.1)--(1.3) with the help
of (1.4) we obtain the following system of equations:
$$
\dfrac{\partial f_1}{\partial t}+v_{x}\dfrac{\partial f_1}{\partial
x}+\nu f_1(x, \mathbf{v}, t)
=\delta(\E_{F}-\E) \big[e E_{x}(x,t)v_{x}+\nu[\E_{F}(x,t)-\E_{F}]\big] ,
\eqno{(1.5)}
$$
$$
\dfrac{\partial E_{x}(x,t)}{\partial x}=\dfrac{8\pi
e}{(2\pi\hbar)^{3}}\int f_1(x,\mathbf{v'}, t)d^3p' 
\eqno {(1.6)}
$$

From the law of preservation of number of particles
$$
\int f_{eq}d\Omega_F=\int f d\Omega_F
$$
we find:
$$
[\E_{F}(x,t)-\E_{F}]\int\delta(\E_{F}-\E)d^3p=\int f_1 d^3p.
\eqno{(1.7)}
$$

From the equation (1.5) it is seen that we should search for
the function $f_1$ in the form proportional to the delta -- function:
$$
f_1= \delta(\E_F-\E) H(x,\mu,t), \qquad \mu=\dfrac{v_x}{v}.
\eqno{(1.8)}
$$

The system of equations (1.5) and (1.6) with the help of (1.7) and (1.8)
can be transformed to the following form:
$$
\dfrac{\partial H}{\partial t}+v_{F}\mu \dfrac{\partial H}{\partial
x}+ \nu H(x,\mu,t)=
{ev_{F}\mu}E_{x}(x,t)+
\dfrac{\nu}{2}\int_{-1}^{1}H(x,\mu',t)d\mu',
$$
$$
\dfrac{\partial E_{x}(x,t)}{\partial x}=\dfrac{16\pi^{2}e m^2
v_{F}}{(2\pi \hbar)^3}\int_{-1}^{1}H(x,\mu',t)d\mu'.
$$

Further we introduce dimensionless functions
$$
e(x_1,t)=\dfrac{E_{x}(x,t)}{E_0},\qquad h(x_1,\mu,t)=
\dfrac{H(x,\mu,t)}{eaE_0},
$$
and pass to dimensionless coordinate $x_1=x/a$. 
We obtain the following system of equations
$$
\dfrac{\partial H}{\partial t_1}+\mu \dfrac{\partial H}{\partial
x_1}+ \nu H(x_1,\mu,t_1)
=\mu e(x_1,t_1)+\dfrac{1}{2}\int_{-1}^{1}H(x_1,\mu',t_1)d\mu',
\eqno{(1.9)}
$$

$$
\dfrac{\partial e(x_1,t_1)}{\partial x_1}=\dfrac{16\pi^2e^2m^2v_Fa^2}
{(2\pi \hbar)^3}\int_{-1}^{1}H(x_1,\mu',t_1)d\mu'.
\eqno{(1.10)}
$$

Here $\omega_{p}$ is the electron (Langmuir) frequency 
of plasma oscillations,
$$
\omega_p^2=\dfrac{4\pi e^2N}{m},
$$
$N$ is the numerical density (concentration), $m$ is the electron mass.

We used the following well-known relation for degenerate plasma for
the conclusion of the equations (1.9) and (1.10)
$$
\Big(\dfrac{v_F m}{\hbar}\Big)^3=3\pi^2 N.
$$

\begin{center}
\item{}\section{Boundary conditions statement}
\end{center}

Let us outline the time variable of the functions $H(x_1,\mu,t_1)$
and $e(x_1,t_1)$, assuming
$$
H(x_1,\mu,t_1)=e^{-i\omega_1t_1}h(x_1,\mu),
\eqno{(2.1)}
$$
$$
e(x_1,t_1)=e^{-i\omega_1t_1}e(x_1).
\eqno{(2.2)}
$$

The system of equations (1.9) and (1.10) in this case will be
transformed to the following form:
$$
\mu\dfrac{\partial h}{\partial x_1}+w_0 h(x_1,\mu)= \mu
e(x_1)+\dfrac{y_0}{2}\int\limits_{-1}^{1}h(x_1,\mu')d\mu',
\eqno{(2.3)}
$$
$$
\dfrac{de(x_1)}{dx_1}=\dfrac{u_p^2}{2}
\int\limits_{-1}^{1}h(x_1,\mu')d\mu'.
\eqno{(2.4)}
$$

Here 
$$
w_0=y_0-ix_0=\dfrac{a}{v_F}(\nu-i\omega),\quad y_0=\dfrac{a\nu}{v_F},
\quad x_0=\dfrac{a\omega}{v_F},\quad
u_p^2=3\Big(\dfrac{a\omega_p}{v_F}\Big)^2.
$$

The constant $u_p$ can be expressed through Debaye radius $r_D$
$$
u_p^2=3\dfrac{a^2}{r_D^2}, \qquad r_D=\dfrac{v_F}{\omega_p}.
$$

Further instead of $x_1$ we write $x$. We rewrite the system
of equations (2.3) and (2.4) in the form:
$$
\mu\dfrac{\partial h}{\partial x}+w_0h(x,\mu)= \mu
e(x)+\dfrac{y_0}{2}\int\limits_{-1}^{1}h(x,\mu')d\mu'.\quad
\eqno{(2.5)}
$$
$$
\dfrac{de(x)}{dx}=\dfrac{u_p^2}{2}\int\limits_{-1}^{1}h(x,\mu')d\mu'.
\eqno{(2.6)}
$$

For electric field in plasma on its   border the boundary 
condition is satisfied
$$
e(-1)=e_s, \qquad e(+1)=e_s,
\eqno{(2.7)}
$$
where $1$ is the dimensionless depth (width) of a semilayer.

Condition of symmetry of boundary conditions (2.7) and the equations (2.5)
and (2.6) mean, that electric field $e(x)$
in the layer possess properties of symmetry
$$
e(x)=e(-x).
\eqno{(2.7a)}
$$

The non-flowing condition for the particle (electric current) flow
through the plasma boundary means that
$$
\int\limits_{-1}^{1}\mu\,h(-1,\mu)\,d\mu=
\int\limits_{-1}^{1}\mu\,h(1,\mu)\,d\mu=0.
\eqno{(2.8)}
$$

In the kinetic theory for the description of the surface properties
the accommodation coefficients are used often. Tangential momentum
and energy accommodation coefficients are the most--used. For the
problem considered the normal electron momentum accommodation under
the scat\-te\-ring on the surface has the most important significance.

Owing to properties of symmetry of electric field and distribution 
function concerning a plane --- middle of a layer --- further
let's enter the coefficient of accommodation of a normal momentum 
through momentim of electron streams
on the bottom surface of the layer.

The normal momentum accommodation coefficient
is defined by the following relation
$$
\alpha_p=\dfrac{P_i-P_r}{P_i-P_s}, \qquad 0\leqslant \alpha_p
\leqslant 1,
\eqno{(2.9)}
$$
where $P_i$ and $P_r$  are are the flows of normal to the surface
momentum of incoming to the boundary and reflected from it electrons,
$$
P_i=\int\limits_{-1}^{0}\mu^2h(-1,\mu)d\mu,
\eqno{(2.10)}
$$
$$
P_r=\int\limits_{0}^{1}\mu^2h(-1,\mu)d\mu,
\eqno{(2.11)}
$$
quantity $P_s$ is the normal momentum flow for electrons
reflected from the surface which are in thermodynamic
equilibrium with the wall,
$$
P_s=\int\limits_{0}^{1}\mu^2h_s(\mu)d\mu,
\eqno{(2.12)}
$$
where the function
$$
h_s(\mu)=A_s,\qquad 0<\mu<1,
$$
is the equilibrium distribution function of the corresponding electrons.
This function is to satisfy the condition
similar to the non-flowing condition
$$
\int\limits_{-1}^{0}\mu h(-1,\mu)d\mu+\int\limits_{0}^{1}\mu
h_s(\mu)d\mu=0.
\eqno{(2.13)}
$$

We are going to consider the relation between the normal momentum
accommodation coefficient $\alpha_p$ and the diffuseness coefficient
$q$ for the case of specular and diffuse boundary conditions which
are written in the following form
$$
h(-1,\mu)=(1-q) h(-1,-\mu)+a_s, \quad 0<\mu<1.
\eqno{(2.14)}
$$

Here $q$ is the diffusivity coefficient ($0\leqslant q\leqslant 1$),
$a_s$ is the quantity determined from the non-flowing condition.

From the non-flowing condition we derive
$$
\int\limits_{-1}^{1}\mu h(-1,\mu)d\mu=\int\limits_{-1}^{0}\mu
h(-1,\mu)d\mu+\int\limits_{0}^{1}\mu h(-1,\mu)d\mu=0.
$$

In the second integral we replace the integrand according to the
right-hand side of the specular--diffuse boundary condition (2.14).
After that, using the obvious change of integration variable, we
obtain that
$$
a_s=-2q\int\limits_{-1}^{0}\mu h(-1,\mu)d\mu.
$$

Let us use the boundary condition (2.13). Using the analogous to the
preceded line of reasoning we get
$$
A_s=-2\int\limits_{-1}^{0}\mu h(-1,\mu)d\mu.
$$

From the two last equations we find that
$$
a_s=qA_s.
\eqno{(2.15)}
$$

Further we find the difference between two flows
$$
P_i-P_r=\int\limits_{-1}^{0}\mu^2
h(-1,\mu)d\mu-\int\limits_{0}^{1}\mu^2 h(-1,\mu)d\mu.
$$

In the second integral we use the boundary condition (2.14) again.
With the help of (2.15) we obtain that
$$
P_i-P_r=q\int\limits_{-1}^{0}\mu^2 h(-1,\mu)d\mu-\int\limits_{0}^{1}\mu^2
a_sd\mu=
$$
$$
=q\int\limits_{-1}^{0}\mu^2 h(-1,\mu)d\mu-q\int\limits_{0}^{1}\mu^2
A_sd\mu=qP_i-qP_s.
$$

Substituting the expressions obtained to the definition of the normal
mo\-men\-tum accommodation coefficient, we have
$$
\alpha_p=\frac{P_i-P_r}{P_i-P_s}=\frac{qP_i-qP_s}{P_i-P_s}=q.
$$

Thus, for specular -- diffuse boundary conditions normal momentum
accom\-mo\-da\-tion coefficient $\alpha_p$
coincides with the diffuseness coefficient $q$.

Equally with the specular -- diffuse boundary conditions another
vari\-ants of boundary conditions are used in kinetic theory as well.

In particular, accommodation boundary conditions are used widely.
They are divided into two types: diffuse -- accommodative and
specular -- accommodative boundary conditions (see \cite{14}).

We consider specular -- accommodative boundary conditions. For
the function $h$ this conditions will be written in the following form
$$
h(-1,\mu)=h(-1,-\mu)+A_0+A_1\mu, \qquad 0<\mu<1.
\eqno{(2.16)}
$$
$$
h(1,\mu)=h(1,-\mu)-A_0+A_1\mu, \qquad -1<\mu<0.
\eqno{(2.17)}
$$

Let's notice, that (2.17) the same appearance, as (2.16) has conditions. 
Really, let's replace $\mu$ on $-\mu$ in the condition (2.17). 
We will receive a condition
$$
h(1,-\mu)=h(1,\mu)-A_0-A_1\mu,\qquad 0<\mu<1,
$$
whence we receive in accuracy a condition (2.16). It means, that both 
conditions  (2.16) and (2.17) it is possible to write down in the form 
of one
$$
h(\pm 1,\mu)=h(\pm 1,-\mu)+A_0+A_1\mu,\qquad 0<\mu<1.
$$

If in (2.16) we assume $A_0=A_1=0$, then specular -- accommodative
boundary conditions pass into pure specular boundary conditions.

Coefficients $A_0$ and $A_1$ can be derived from the non-flowing
condition and the definition of the normal electron momentum
accommodation coefficient.

The problem statement is completed. Now the problem consists in
finding of such solution of the system of equations (2.5) and (2.6),
which satisfies the boundary conditions  (2.7) and (2.16). Further, with
the use of the solution of the problem, it is required to built the
profiles of the distribution function of the electrons moving to the
plasma surface, and profile of the electric field.

\begin{center}
\item{}\section{The relation between flows and boundary conditions}
\end{center}

First of all let us find expression which relates the constants
$A_0,\;A_1$ from the boundary condition (2.16). To carry this out
we will use the condition of non-flowing (2.12) of the particle flow
through the plasma boundary, which
we will write as a sum of two flows
$$
N_0\equiv \int\limits_{0}^{1}\mu h(-1,\mu)d\mu+\int\limits_{-1}^{0}
\mu h(-1,\mu)d\mu=0.
$$

After evident substitution of the variable in the second
integral we obtain
$$
N_0\equiv \int\limits_{0}^{1} \mu\Big[h(-1,\mu)-h(-1,-\mu)\Big]d\mu=0.
$$

Taking into account the relation  (2.16), we obtain that
$
A_0=-\frac{2}{3}A_1.
$

With the help of this relation we can rewrite the condition (2.16)
in the following form
$$
h(-1,\mu)-h(-1,-\mu)=A_1(\mu-\dfrac{2}{3}), \qquad 0<\mu<1,
\eqno{(3.1)}
$$
or, that is equivalent,
$$
h(1,\mu)-h(1,-\mu)=A_1(\mu-\dfrac{2}{3}), \qquad 0<\mu<1,
\eqno{(3.1')}
$$

We consider the momentum flow of the electrons which are moving to
the boundary. According to (3.1) we have
$$
P_i=P_r-\dfrac{1}{36}A_1.
\eqno{(3.2)}
$$

It is easy to see further that
$$
P_s=\frac{A_s}{3}. 
\eqno{(3.3)}
$$

With the help of the formulas (3.2) and (3.3) we will rewrite the
definition of the accommodation coefficient (2.9) in the form
$$
\alpha_pP_r-\alpha_p\dfrac{A_s}{3}+\dfrac{A_1}{36}(1-\alpha_p)=0.
\eqno{(3.4)}
$$

Let us consider the condition (2.13). We rewrite it in the following form
$$
\dfrac{A_s}{2}+\int\limits_{-1}^{0}\mu h(-1,\mu)d\mu=0.
$$

From this condition we obtain
$$
A_s=-2\int\limits_{-1}^{0}\mu h(-1,\mu)d\mu=2\int\limits_{0}^{1}
\mu h(-1,-\mu)d\mu.
$$

Using the condition (3.1), we then get
$$
A_s=2\int\limits_{0}^{1}\mu h(-1,\mu)d\mu.
\eqno{(3.5)}
$$

Now with the help of the second equality from (2.11) and (3.5) we
rewrite the relation (3.4) in the integral form
$$
\alpha_p\int\limits_{0}^{1}\Big(\mu^2-\dfrac{2}{3}\mu\Big)h(-1,\mu)d\mu=
-\dfrac{1}{36}(1-\alpha_p)A_1.
\eqno{(3.6)}
$$

Now the boundary problem consists of the equations  (2.5) and (2.6)
and boundary conditions (2.7), (3.1) and (3.6).

\begin{center}
\item{}\section{Separation of variables and characteristic system}
\end{center}

Application of the general Fourier method of the separation of
variables in several steps results in the following substitution \cite{Case}
$$
h_\eta(x,\mu)=\exp(-\dfrac{w_0x}{\eta})\Phi(\eta,\mu)+
\exp(\dfrac{z_0x}{\eta})\Phi(-\eta,\mu),
\eqno{(4.1)}
$$
$$
e_\eta(x)=\Big[\exp(-\dfrac{w_0x}{\eta})+\exp(\dfrac{w_0x}{\eta})\Big]
E(\eta),
\eqno{(4.2)}
$$
where $\eta$ is the spectrum parameter or the parameter of
separation, which is complex in general.

We substitute the equalities (4.1) and (4.2) into the equations 
(2.5) and (2.6).
We obtain the following characteristic system of equations
$$
w_0(\eta-\mu)\Phi(\eta,\mu)=\eta\mu E(\eta)+\dfrac{\eta}{2}
\int\limits_{-1}^{1}\Phi(\eta,\mu')d\mu',
\eqno{(4.3)}
$$
$$
w_0(\eta+\mu)\Phi(-\eta,\mu)=\eta\mu E(\eta)+\dfrac{\eta}{2}
\int\limits_{-1}^{1}\Phi(-\eta,\mu')d\mu',
\eqno{(4.4)}
$$
$$
-\dfrac{w_0}{\eta}E(\eta)=u_p^2\cdot \dfrac{1}{2}
\int\limits_{-1}^{1}\Phi(\eta,\mu')d\mu',
\eqno{(4.5)}
$$
$$
\dfrac{w_0}{\eta}E(\eta)=u_p^2\cdot
\dfrac{1}{2}\int\limits_{-1}^{1}\Phi(-\eta,\mu')d\mu'.
\eqno{(4.6)}
$$

From the equations (4.5) and (4.6) we obtain
$$
\int\limits_{-1}^{1}\Phi(\eta,\mu)d\mu=-
\int\limits_{-1}^{1}\Phi(-\eta,\mu)d\mu.
\eqno{(4.7)}
$$

Let us introduce the designations
$$
n(\eta)=\int\limits_{-1}^{1}\Phi(\eta,\mu)d\mu.
\eqno{(4.8)}
$$

From the equation (4.5) we find, that
$$
E(\eta)=-\dfrac{u_p^2}{2w_0}\eta n(\eta),
\eqno{(4.9)}
$$
whence
$$
n(\eta)=-2\dfrac{w_0}{u_p^2}\cdot\dfrac{E(\eta)}{\eta}.
$$

By means of equalities (4.7) -- (4.9) we will copy the equations (4.3) and
(4.4)
$$
(\eta-\mu)\Phi(\eta,\mu)=\dfrac{E(\eta)}{w_0}(\mu\eta-\eta_1^2),
\eqno{(4.10)}
$$
$$
(\eta+\mu)\Phi(-\eta,\mu)=\dfrac{E(\eta)}{w_0}(\mu\eta+\eta_1^2).
\eqno{(4.11)}
$$

Here
$$
\eta_1^2=\dfrac{v_0w_0}{u_p^2}=\dfrac{\nu(\nu-i\omega)}{3\omega_p^2}=
\dfrac{\varepsilon^2 z_0}{3}, \qquad \varepsilon=\dfrac{\nu}{\omega_p},
\qquad z_0=1-i\dfrac{\omega}{\nu}=1-i\dfrac{\Omega}{\varepsilon},\qquad
\Omega=\dfrac{\omega}{\omega_p}.
$$

Solution of the system (4.10) and (4.11) depends essentially on the
condition whether the spectrum parameter $\eta$ belongs to the interval 
$-1<\eta<1$. In connection with this the interval $-1<\eta<1$ we will 
call as continuous spectrum of the characteristic system.

Let the parameter $\eta\in (-1,1)$. Then from the equations (4.10) and 
(4.11) in the class of general functions we will find eigenfunction
corresponding to the continuous spectrum
$$
\Phi(\eta,\mu)=\dfrac{E(\eta)}{z_0}P\dfrac{\mu\eta-\eta_1^2}{\eta-\mu}+
g_1(\eta)\delta(\eta-\mu),
\eqno{(4.12)}
$$
$$
\Phi(-\eta,\mu)=\dfrac{E(\eta)}{z_0}P\dfrac{\mu\eta+\eta_1^2}{\eta+\mu}+
g_2(\eta)\delta(\eta-\mu).
\eqno{(4.13)}
$$

In these equations (4.12) and (4.13) $\delta(x)$ is 
the delta--function of Dirac,
the symbol $Px^{-1}$ means the principal value of the integral under
integrating of the expression $x^{-1}$.

Substituting now (4.12) and (4.13) in the equations (4.5) and (4.6),
we receive the equations from which we obtain
$$
g_1(\eta)=-2\dfrac{w_0}{u_p^2}\dfrac{\lambda(\eta)}{\eta}E(\eta),\qquad
g_2(\eta)=-g_1(\eta)=2\dfrac{w_0}{u_p^2}\dfrac{\lambda(\eta)}{\eta}E(\eta).
\eqno{(4.14)}
$$

Here dispersion function is entered
$$
\lambda(z)=1-\dfrac{z}{2c}\int\limits_{-1}^{1}\dfrac{\mu z-\eta_1^2}
{\mu-z}d\mu,
\eqno{(4.15)}
$$
where
$$
c=\dfrac{w_0^2}{u_p^2}=\dfrac{\varepsilon^2z_0^2}{3}=z_0\eta_1^2.
$$

Functions (4.12) and (4.13) are  called eigen functions of the continuous
spectrum, since the spectrum parameter $\eta$ fills out the
continuum $(-1,+1)$ compactly. The eigen solutions of the given
problem can be found from the equalities (4.1) and (4.2).

Substituting relations (4.14) in (4.12) and (4.13), we will present
last expressions in the following form
$$
\Phi(\eta,\mu)=\dfrac{E(\eta)}{w_0}
\Big[P\dfrac{\mu\eta-\eta_1^2}{\eta-\mu}-2c\dfrac{\lambda(\eta)}
{\eta}\delta(\eta-\mu)\Big],
$$
$$
\Phi(-\eta,\mu)=\dfrac{E(\eta)}{w_0}
\Big[P\dfrac{\mu\eta+\eta_1^2}{\eta+\mu}+2c\dfrac{\lambda(\eta)}
{\eta}\delta(\eta+\mu)\Big],
$$
or
$$
\Phi(\eta,\mu)=\dfrac{E(\eta)}{w_0}F(\eta,\mu),\qquad
\Phi(-\eta,\mu)=\dfrac{E(\eta)}{w_0}F(-\eta,\mu),
$$
where
$$
F(\eta,\mu)=P\dfrac{\mu\eta-\eta_1^2}{\eta-\mu}-2c\dfrac{\lambda(\eta)}
{\eta}\delta(\eta-\mu),
$$
$$
F(-\eta,\mu)=P\dfrac{\mu\eta+\eta_1^2}{\eta+\mu}+2c\dfrac{\lambda(\eta)}
{\eta}\delta(\eta+\mu).
$$

It will be necessary for us the following relation of symmetry
$
F(-\eta,-\mu)=-F(\eta,\mu).
$

Let us notice, that eigen function $F(\eta,\mu)$ satisfies
to following condition of normalization
$$
\int\limits_{-1}^{1}F(\pm \eta,\mu)d\mu=\mp 2cP\dfrac{1}{\eta}.
$$

So, eigen function of a continuous spectrum is constructed and
it is defined by equality
$$
h_\eta(x,\mu)=\Big[\exp\Big(-\dfrac{xw_0}{\eta}\Big)F(\eta,\mu)+
\exp\Big(\dfrac{xw_0}{\eta}\Big)F(-\eta,\mu)\Big]\dfrac{E(\eta)}{w_0},
$$
or, in explicit form,
$$
h_\eta(x,\mu)=\Bigg\{\Big[\exp\Big(-\dfrac{xw_0}{\eta}\Big)
\dfrac{\mu\eta-\eta_1^2}{\eta-\mu}+
\exp\Big(\dfrac{xw_0}{\eta}\Big)
\dfrac{\mu\eta+\eta_1^2}{\eta+\mu}\Big]-
$$
$$
-2c\dfrac{\lambda(\eta)}{\eta}\Big[\exp\Big(-\dfrac{xw_0}{\eta}\Big)
\delta(\eta-\mu)-\exp\Big(\dfrac{xw_0}{\eta}\Big)\delta(\eta+\mu)\Big]
\Bigg\}
\dfrac{E(\eta)}{w_0}.
$$

Let us replace exponents by hyperbolic functions 
$$
\exp\Big(\dfrac{xw_0}{\eta}\Big)=
\ch\dfrac{xw_0}{\eta}+\sh\dfrac{xw_0}{\eta},\qquad
\exp\Big(-\dfrac{xw_0}{\eta}\Big)=\ch\dfrac{xw_0}{\eta}-
\sh\dfrac{xw_0}{\eta}
$$
and also we will transform both square brackets from the 
previous expression.  As a result we receive, that 
for the first square bracket it is had
$$
\exp\Big(-\dfrac{xw_0}{\eta}\Big)
\dfrac{\mu\eta-\eta_1^2}{\eta-\mu}+
\exp\Big(\dfrac{xw_0}{\eta}\Big)
\dfrac{\mu\eta+\eta_1^2}{\eta+\mu}=
$$
$$
=\exp\Big(-\dfrac{xw_0}{\eta}\Big)
\dfrac{(\mu\eta-\eta_1^2)(\eta+\mu)}{\eta^2-\mu^2}+
\exp\Big(\dfrac{xw_0}{\eta}\Big)
\dfrac{(\mu\eta+\eta_1^2)(\eta-\mu)}{\eta^2-\mu^2}=
$$
$$
=2\Big[\sh\dfrac{xw_0}{\eta}P\dfrac{\eta(\mu^2-\eta_1^2)}{\mu^2-\eta}-
\ch\dfrac{xw_0}{\eta}P\dfrac{\mu(\eta^2-\eta_1^2)}{\mu^2-\eta^2}\Big].
$$

For the second square bracket we have
$$
\exp\Big(-\dfrac{xw_0}{\eta}\Big)
\delta(\eta-\mu)-\exp\Big(\dfrac{xw_0}{\eta}\Big)\delta(\eta+\mu)=
$$
$$
=-\Big[\delta(\eta-\mu)+\delta(\eta+\mu)\Big]\sh\dfrac{xw_0}{\eta}+
\Big[\delta(\eta-\mu)-\delta(\eta-\mu)\Big]\ch\dfrac{xw_0}{\eta}.
$$

As a result we receive, that
$$
h_\eta(x,\mu)=\dfrac{2E(\eta)}{w_0}
\Bigg\{\ch\dfrac{xw_0}{\eta}\Bigg[ P\Big(
\dfrac{\mu\eta-\eta_1^2}{\eta-\mu}+\dfrac{\mu\eta+\eta_1^2}
{\eta+\mu}\Big)+
$$
$$
+2c\dfrac{\lambda(\eta)}{\eta}(-\delta(\eta-\mu)+
\delta(\eta+\mu))\Bigg]+\sh\dfrac{xw_0}{\eta}\Bigg[P
\Big(-\dfrac{\mu\eta-\eta_1^2}{\eta-\mu}+\dfrac{\mu\eta+\eta_1}
{\eta+\mu}\Big)+
$$
$$
+2c\dfrac{\lambda(\eta)}{\eta}(\delta(\eta-\mu)+\delta(\eta+\mu))\Bigg]
\Bigg\}.
$$

We will designate further
$$
\varphi(\eta,\mu)=F(\eta,\mu)+F(-\eta,\mu)=
P\Big(\dfrac{\mu\eta-\eta_1^2}{\eta-\mu}+\dfrac{\mu\eta+\eta_1^2}
{\eta+\mu}\Big)=
2P\dfrac{\mu(\eta^2-\eta_1^2)}{\eta^2-\mu^2},
$$
and
$$
\psi(\eta,\mu)=-F(\eta,\mu)+F(-\eta,\mu)=P
\Big(-\dfrac{\mu\eta-\eta_1^2}{\eta-\mu}+\dfrac{\mu\eta+\eta_1}
{\eta+\mu}\Big)=2P\dfrac{\eta(\mu^2-\eta_1^2)}{\eta^2-\mu^2}.
$$

Thus, eigen function of a continuous spectrum it is possible
to present in the form of a linear combination of a hyperbolic sine and
kosine
$$
h_\eta(x,\mu)=\dfrac{2E(\eta)}{w_0}\Bigg\{\ch\dfrac{xw_0}{\eta}\Big[
P\dfrac{\mu(\eta^2-\eta_1^2)}{\eta^2-\mu^2}-c\dfrac{\lambda(\eta)}{\eta}
\big(\delta(\eta-\mu)-\delta(\eta+\mu)\big)\Big]-
$$
$$
-\sh\dfrac{xw_0}{\eta}\Big[P\dfrac{\eta(\mu^2-\eta_1^2)}{\eta^2-\mu^2}-
c\dfrac{\lambda(\eta)}{\eta}
\big(\delta(\eta-\mu)+\delta(\eta+\mu)\big)\Big]\Bigg\}.
$$

Let us notice, that eigen functions of a continuous spectrum it is possible
to present and in such form
$$
h_\eta(x,\mu)=\dfrac{E(\eta)}{w_0}\Bigg[\ch\dfrac{xw_0}{\eta}
\Big(F(\eta,\mu)+F(-\eta,\mu)\Big)
+\sh\dfrac{xw_0}{\eta}\Big(-F(\eta,\mu)+F(-\eta,\mu)\Big)\Bigg].
\eqno{(4.16)}
$$

The dispersion function $\lambda(z)$ we express in the terms of
the Case dispersion function \cite{Case}
$$
\lambda(z)=1-\dfrac{1}{z_0}+\dfrac{1}{z_0}\Big(1-
\dfrac{z^2}{\eta_1^2}\Big)\lambda_C(z),
$$
where
$$
\lambda_C(z)=1+\dfrac{z}{2}\int\limits_{-1}^{1}\dfrac{d\tau}{\tau-z}=
\dfrac{1}{2}\int\limits_{-1}^{1}\dfrac{\tau\,d\tau}{\tau-z}
$$
is the Case dispersion function.

In the complex plane dispersion Case function is calculated
through the logarithm
$$
\lambda_C(z)=1+\dfrac{z}{2}\ln\dfrac{z-1}{z+1}, \qquad
z\in \mathbb{C}\setminus [-1,1],
$$
and on cut by formula
$$
\lambda_0(\eta)=1+\dfrac{\eta}{2}\ln\dfrac{1-\eta}{1+\eta}, \qquad
\eta\in (-1,1).
$$

The boundary values of the dispersion function from above and below
the cut (interval $(-1,1)$) we define in the following way
$$
\lambda^{\pm}(\mu)=\lim\limits_{\varepsilon\to 0, \varepsilon>0}
\lambda(\mu\pm i \varepsilon), \qquad \mu\in (-1,1).
$$

The boundary values of the dispersion function from above and below
the cut are calculated according to the Sokhotzky formulas
$$
\lambda^{\pm}(\mu)=\lambda(\mu)\pm \dfrac{i \pi\mu}
{2\eta_1^2z_0}(\eta_1^2-\mu^2),\quad -1<\mu<1,
$$
from where
$$
\lambda^+(\mu)-\lambda^-(\mu)=\dfrac{i \pi}{\eta_1^2z_0}
\,\mu(\eta_1^2-\mu^2),\qquad
\dfrac{\lambda^+(\mu)+\lambda^-(\mu)}{2}=\lambda(\mu),\quad-1<\mu<1,
$$
where
$$
\lambda(\mu)=1+\dfrac{\mu}{2\eta_1^2z_0} \int\limits_{-1}^{1}
\dfrac{\eta_1^2-\eta^2}{\eta-\mu}\,d\eta,
$$
and the integral in this equality is understood as singular in terms
of the principal value by Cauchy. Besides that, the function
$\lambda(\mu)$ can be represented in the following form
$$
\lambda(\mu)=1-\dfrac{1}{z_0}+
\dfrac{1}{z_0}\Big(1-\dfrac{\mu^2}{\eta_1^2}\Big)\lambda_0(\mu),\qquad
\lambda_0(\mu)=1+\dfrac{\mu}{2}\ln\dfrac{1-\mu}{1+\mu}.
$$

\begin{center}
\item{}\section{Eigen functions of discrete spectrum and plasma waves}
\end{center}

According to the definition, the discrete spectrum of the
characteristic equation is a set of zeroes of the dispersion
equation
$$
\dfrac{\lambda(z)}{z}=0. 
\eqno{(5.1)}
$$

We start to search zeroes of the equation (5.1). Let us take Laurent
series of the dispersion function
$$
\lambda(z)=\lambda_\infty+\dfrac{\lambda_2}{z^2}+
\dfrac{\lambda_4}{z^4}+\cdots,\qquad |z|>1.
\eqno{(5.2)}
$$

Here
$$
\lambda_\infty \equiv\lambda(\infty)=
1-\dfrac{1}{z_0}+\dfrac{1}{3z_0\eta_1^2},\quad
\lambda_2=-\dfrac{1}{z_0}\Big(\dfrac{1}{3}-\dfrac{1}{5\eta_1^2}\Big),\quad
\lambda_4=-\dfrac{1}{z_0}\Big(\dfrac{1}{5}-\dfrac{1}{7\eta_1^2}\Big).
$$

We express these parameters through the parameters $\gamma$ and
$\varepsilon$
$$
\lambda_\infty \equiv\lambda(\infty)=
\dfrac{2(\Omega-1)+i\varepsilon+(\Omega-1)(\Omega-1+i\varepsilon)}
{(\Omega+i\varepsilon)^2},
$$
$$
\lambda_2=-\dfrac{9+5i\varepsilon(\Omega+i\varepsilon)}{15(\Omega+
i\varepsilon)^2},\qquad
\lambda_4=-\dfrac{15+7i\varepsilon(\Omega+i\varepsilon)}{35(\Omega+
i\varepsilon)^2}.
$$

It is easy seen that the dispersion function (4.9) in collisional plasma
(i.e. when $\varepsilon>0$) in the infinity has the value which doesn't
equal to zero: $\lambda_\infty=\lambda(\infty)\ne 0$.

Hence, the dispersion equation has infinity as a zero $\eta_i=\infty$,
to which the discrete eigensolutions of the given system correspond
$$
h_\infty(x,\mu)=\dfrac{\mu}{w_0},\qquad\;e_\infty(x)=1.
$$

This solution is naturally called as mode of Drude. It describes the
volume conductivity of metal, considered by Drude
(see, for example, \cite{16}).

Let us consider the question of the plasma mode existence in
details. We find finite complex zeroes of the dispersion function.
We use the principle of argument. We take the contour (see Fig. 1)
$\Gamma_\varepsilon^+=\Gamma_R\cup\gamma_\varepsilon$,
$
\Gamma_R=\{z: |z|=R,\quad\;R=1/\varepsilon, \quad\varepsilon>0\},
$
which is passed in the positive direction the cut $[-1,+1]$, 
and which bounds the biconnected domain $D_R$. 

Let us notice, that dispersion function in area $D_R$ has no poles.
Then owing to the principle of argument \cite {17} zeroes number  
$N$ in area $D_R $ it is equal
$$
N=\dfrac{1}{2\pi
i}\oint\limits_{\Gamma_\varepsilon}d\,\ln\lambda(z).
$$

\begin{figure}
\begin{center}
\includegraphics{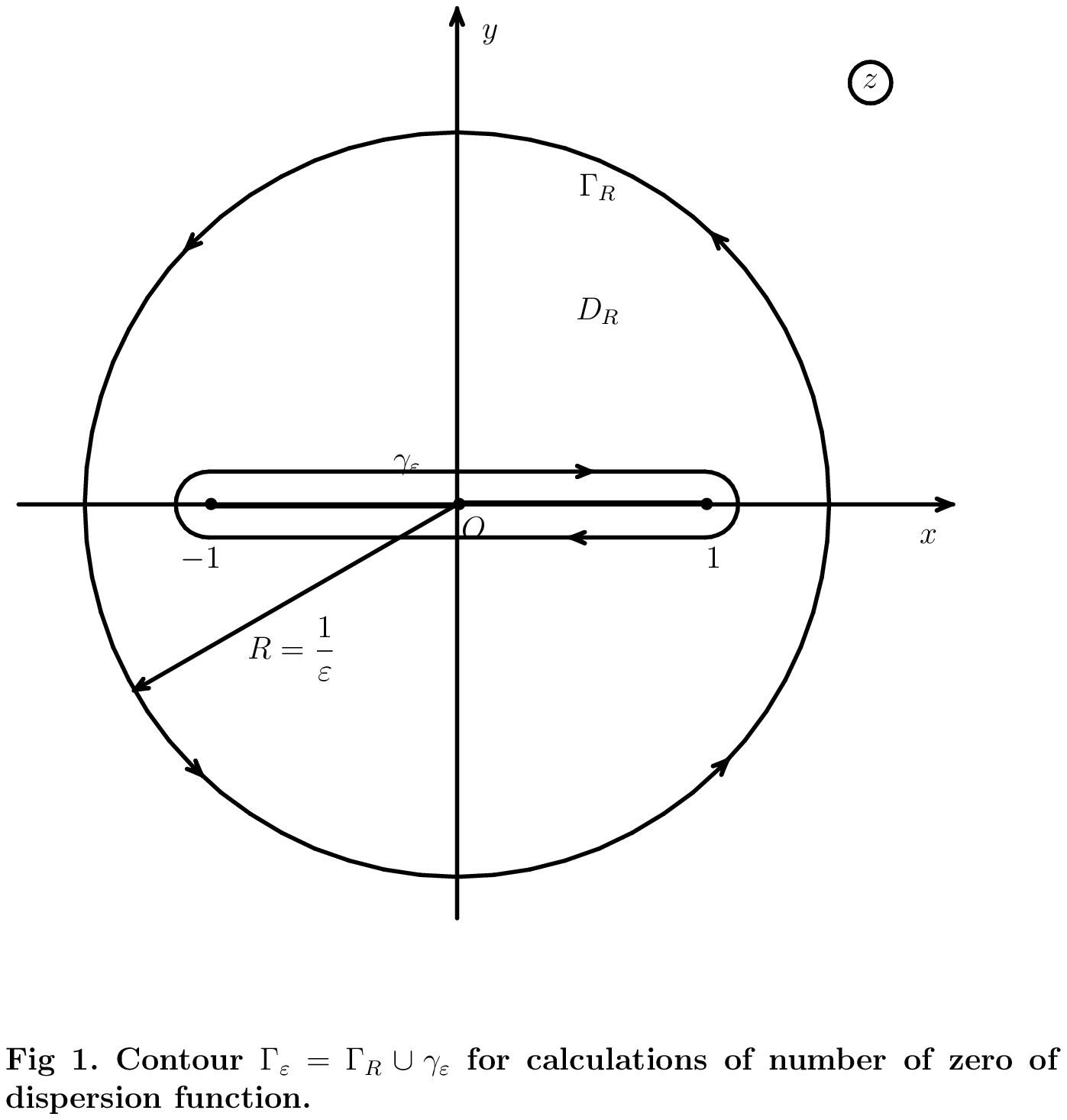}
\end{center}
\begin{center}
{\bf Fig. 1.}
\end{center}
\end{figure}

Considering the limit in this equality when $\varepsilon\to 0$ and
taking into account that the dispersion function is analytic in the
neighbourhood of the infinity, we obtain that
$$
N=\dfrac{1}{2\pi i}\int\limits_{-1}^{1}d\,\ln
\lambda^+(\tau)-\dfrac{1}{2\pi i}\int\limits_{-1}^{1}d\,\ln
\lambda^-(\tau)=\dfrac{1}{2\pi i}\int\limits_{-1}^{1}d\,\ln
\dfrac{\lambda^+(\mu)}{\lambda^-(\mu)}.
$$

So, we have received, that
$$
N=\dfrac{1}{2\pi i}\int\limits_{-1}^{1}d\,\ln
\dfrac{\lambda^+(\tau)}{\lambda^-(\tau)}.
$$

We divide this integral into two integrals by segments $[-1,0]$
and $[0,1]$. In the first integral by the segment $[-1,0]$ we
carry out replacement of variable
$ \tau \rightarrow -\tau $. Taking into account that
$\lambda^+(-\tau)=\lambda^-(\tau)$, we obtain that
$$
N=\dfrac{1}{2\pi i}\int\limits_{-1}^{1}d\,\ln
\dfrac{\lambda^+(\tau)}{\lambda^-(\tau)}= \dfrac{1}{\pi
i}\int\limits_{0}^{1}d\,\ln
\dfrac{\lambda^+(\tau)}{\lambda^-(\tau)}=\dfrac{1}{\pi}\arg
\dfrac{\lambda^+(\tau)}{\lambda^-(\tau)}\Bigg|_0^1.
\eqno{(5.3)}
$$

Here under symbol $\arg G(\tau)=\arg \dfrac{\lambda^+(\tau)}
{\lambda^-(\tau)}$ we understand the regular branch of the argument,
fixed in zero with the condition: $\arg G(0)=0$.

We consider the curve $\Lambda_G=\Lambda_G(\gamma,\varepsilon):
\;z=G(\tau),\;0\leqslant \tau\leqslant+1$,
where
$$
G(\tau)=\dfrac{\lambda^+(\tau)}{\lambda^-(\tau)}.
$$

It is obvious that $G(0)=1,\; \lim\limits_{\tau\to +1}G(\tau)=1$.
Consequently, according to (5.3), the number of values $N$ equals
to doubled number of turns of the curve $\gamma$ around the point
of origin, i.e.
$$
N=2\varkappa(G),
\eqno{(5.3')}
$$
where $\varkappa(G)={\rm Ind}_{[0,+1]}G(\tau)$
is the index of the function $G(\tau)$.

Thus, the number of zeroes of the dispersion function, which are
situated in complex plane outside of the segment $[-1,1]$ of the
real axis, equals to doubled index of the function $G(\tau)$,
calculated on the "semi-segment"\, $[0,+1]$.

Let us single real and imaginary parts of the function $G(\mu)$ out.
At first, we represent the function $G(\mu)$ in the form
$$
G(\mu)=\dfrac{(z_0-1)\eta_1^2+(\eta_1^2-\mu^2)\lambda_0(\mu)+
is(\mu)(\eta_1^2-\mu^2)}{(z_0-1)\eta_1^2+
(\eta_1^2-\mu^2)\lambda_0(\mu)- is(\mu)(\eta_1^2-\mu^2)}.
$$
where
$$
s(\mu)=\dfrac{\pi}{2}\mu, \qquad
\lambda(\mu)=1-\dfrac{1}{z_0}+
\dfrac{1}{z_0}\Big(1-\dfrac{\mu^2}{\eta_1^2}\Big)\lambda_0(\mu),
$$
and
$$
\lambda_0(\mu)=1+\dfrac{\mu}{2}\ln\dfrac{1-\mu}{1+\mu}
$$
is the dispersion function of Case, calculated on the cut 
(i.e., in the interval $(-1,1)$).

Taking into account that
$$
z_0-1=-i\dfrac{\omega}{\nu}=-i\dfrac{\Omega}{\varepsilon}, \qquad
\eta_1^2=\dfrac{\varepsilon z_0}{3}=\dfrac{\varepsilon^2}{3}-
i\dfrac{\varepsilon \Omega}{3},\qquad
(z_0-1)\eta_1^2=-\dfrac{\Omega^2}{3}-i\dfrac{\varepsilon \Omega}
{3},
$$
we obtain
$$
G(\mu)=\dfrac{P^-(\mu)+iQ^-(\mu)}{P^+(\mu)+iQ^+(\mu)},
$$
where
$$
P^{\pm}(\mu)=\Omega^2-\lambda_0(\mu)(\varepsilon^2-3\mu^2)\pm
\varepsilon  \Omega s(\mu),
$$
$$
Q^{\pm}(\mu)=\varepsilon \Omega(1+\lambda_0(\mu))\pm
s(\mu)(\varepsilon^2-3\mu^2).
$$

Now we can easily separate real and imaginary parts of the function
$G(\mu)$  
$$
G(\mu)=\dfrac{g_1(\mu)}{g(\mu)}+i\dfrac{g_2(\mu)}{g(\mu)}.
$$

Here
$$
g(\mu)=[P^+(\mu)]^2+[Q^+(\mu)]^2=$$$$+[\Omega^2+\lambda_0(3\mu^2-
\varepsilon^2)-\varepsilon \Omega s]^2+
[\varepsilon \Omega(1+\lambda_0)- s(3\mu^2-\varepsilon^2)]^2,
$$
$$
g_1(\mu)=P^+(\mu)P^-(\mu)+Q^+(\mu)Q^-(\mu)=$$$$=[\Omega^2+
\lambda_0(3\mu^2-\varepsilon^2)]^2-
\varepsilon^2 \Omega^2[s^2-(1+\lambda_0)^2]-
(3\mu^2-\varepsilon^2)^2s^2,
$$
$$
g_2(\mu)=P^+(\mu)Q^-(\mu)-P^-(\mu)Q^+(\mu)=$$$$+
2s[\Omega^2(3\mu^2-\varepsilon^2)+
\lambda_0(3\mu^2-\varepsilon^2)^2+\varepsilon^2 \Omega^2(1+\lambda_0)],
$$

We consider (see Fig. 2) the curve $L$, which is defined in
implicit form by the following parametric equations
$$
L=\big\{(\Omega,\varepsilon): \qquad g_1(\mu;\Omega,\varepsilon)=0,\;\quad
g_2(\mu;\Omega,\varepsilon)=0, \quad 0\leqslant \mu \leqslant 1\big\},
$$
and which lays in the plane of the parameters of the problem
$(\gamma,\varepsilon)$,
and when passing through this curve the index of the function $G(\mu)$
at the positive "semi-segment"\, $[0,1]$ changes stepwise.

From the equarion $g_2=0$ we find
$$
\Omega^2=-\dfrac{\lambda_0(\mu)(3\mu^2-\varepsilon^2)}
{3\mu^2+\varepsilon^2\lambda_0(\mu)}.
\eqno{(5.4)}
$$

Now from the equation $g_1=0$ with the help of (5.4) we find that
$$
\varepsilon=\sqrt{L_2(\mu)},
\eqno{(5.5)}
$$
where
$$
L_2(\mu)=-\dfrac{3\mu^2s^2(\mu)}{\lambda_0(\mu)
[s^2(\mu)+(1+\lambda_0(\mu))^2]}.
$$

Substituting (5.5) into (5.4), we obtain
$$
\Omega=+\sqrt{L_1(\mu)}, 
\eqno{(5.6)}
$$
where
$$
L_1(\mu)=-\dfrac{3\mu^2[s^2(\mu)+
\lambda_0(\mu)(1+\lambda_0(\mu))]^2}{\lambda_0(\mu)[s^2(\mu)+
(1+\lambda_0(\mu))^2]}.
$$

Functions (5.5) and (5.6) determine the curve $L$ which is the
border if the domain $D^+$ (we designate the external area to the
domain as $D^-$) in explicit parametrical form (see Fig. 2). As in
the work \cite{18} we can prove that if $(\gamma,\varepsilon)\in
D^+$, then $\varkappa(G)={\rm Ind}_{[0,+1]} G(\mu)=1$ (the curve $L$
encircles the point of origin once), and if $(\gamma,\varepsilon)\in
D^-$, then $\varkappa(G)={\rm Ind}_{[0,+1]} G(\mu)=0$ (the curve $L$
doesn't encircle the point of origin).

We note, that in the work \cite{18} 
the method of analysis of boundary regime when
$(\gamma,\varepsilon)\in L$ was developed.

From the expression (3.2) one can see that the number of zeroes of
the dispersion function equals to two if $(\gamma,\varepsilon)\in
D^+$, and equals to zero if $(\gamma,\varepsilon)\in D^-$.
\begin{figure}
\begin{center}
\includegraphics[width=16cm, height=10cm]{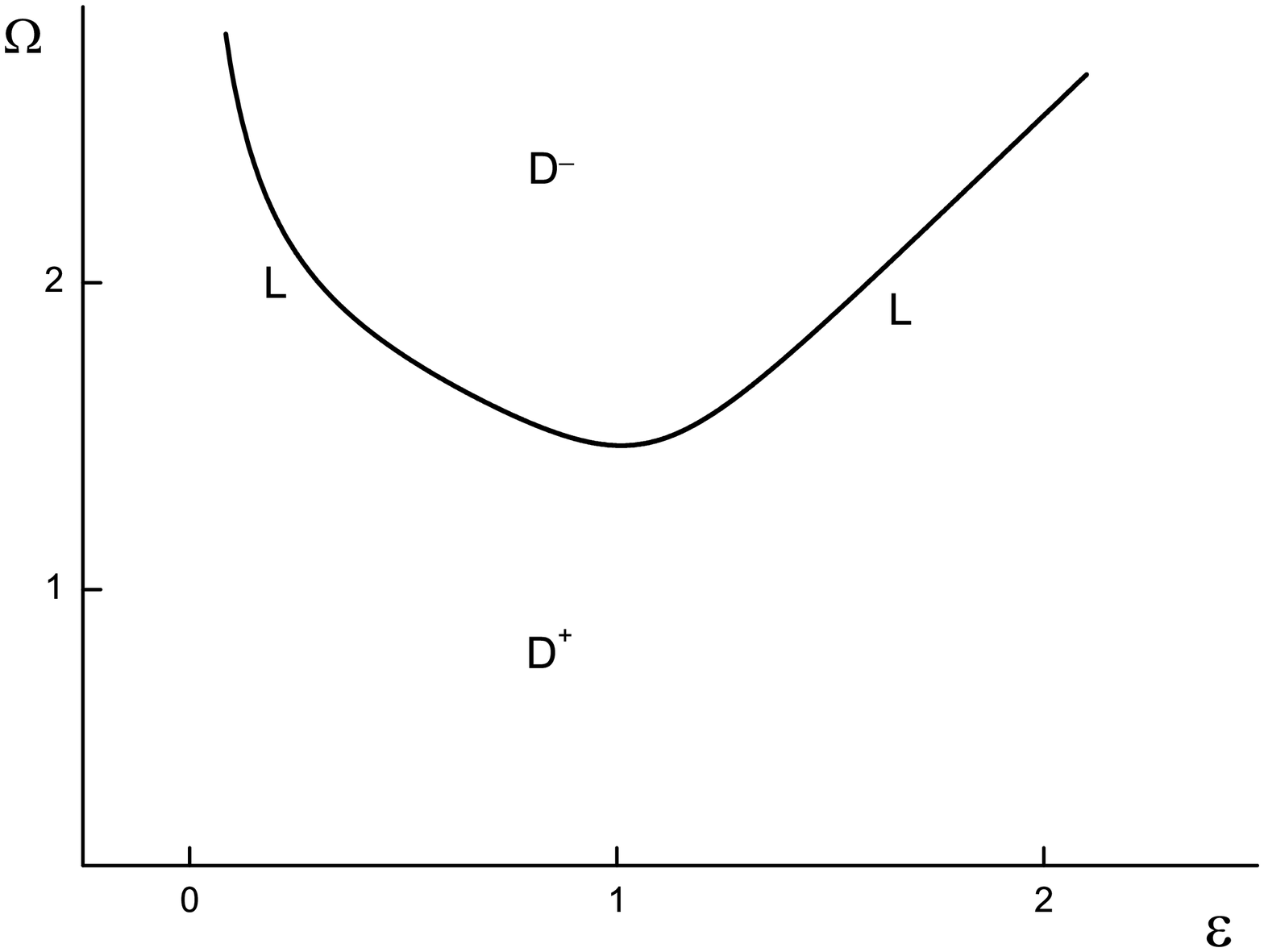}
\end{center}
\begin{center}
{\bf Fig. 2.}
\end{center}
\end{figure}

Since the dispersion function is even its zeroes differ from each
other by sign. We designate these zeroes as following $\pm \eta_0$,
by $\eta_0$ we take the zero which satisfies the condition $\Re
\eta_0>0$. 

The following solution corresponds to the zero $\pm\eta_0$
$$
h_{\eta_0}(x,\mu)=\ch\dfrac{xw_0}{\eta_0}[F(\eta_0,\mu)+F(-\eta_0,\mu)]
+\sh\dfrac{xw_0}{\eta_0}[-F(\eta_0,\mu)+F(-\eta_0,\mu)],
\eqno{(5.7)}
$$
$$
e_{\eta_0}(x)=2\ch\dfrac{w_0x}{\eta_0}.
\eqno{(5.8)}
$$

Here
$$
F(\eta_0,\mu)=\dfrac{\eta_0\mu-\eta_1^2}{\eta_0-\mu},\qquad
F(-\eta_0,\mu)=\dfrac{\eta_0\mu+\eta_1^2}{\eta_0+\mu}.
$$

It is easy to see, that function $h_{\eta_0}(x,\mu) $ is even
on $\eta_0$
$$
h_{\eta_0}(x,\mu)=h_{-\eta_0}(x,\mu).
$$

Function $h_{\eta_0}(x,\mu) $ we will present in the explicit form
$$
h_{\eta_0}(x,\mu)=\ch\dfrac{xw_0}{\eta_0}\Bigg[
\dfrac{\eta_0\mu-\eta_1^2}{\eta_0-\mu}+
\dfrac{\eta_0\mu+\eta_1^2}{\eta_0+\mu}\Bigg]+
+\sh\dfrac{xw_0}{\eta_0}\Bigg[-
\dfrac{\eta_0\mu-\eta_1^2}{\eta_0-\mu}+
\dfrac{\eta_0\mu+\eta_1^2}{\eta_0+\mu}\Bigg],
\eqno{(5.9)}
$$
or
$$
h_{\eta_0}(x,\mu)=\ch\dfrac{xw_0}{\eta_0}\varphi(\eta_0,\mu)
+\sh\dfrac{xw_0}{\eta_0}\psi(\eta_0,\mu).
\eqno{(5.9')}
$$

Here
$$
\varphi(\eta_0,\mu)=F(\eta_0,\mu)+F(-\eta_0,\mu)
=\dfrac{\eta_0\mu-\eta_1^2}{\eta_0-\mu}+
\dfrac{\eta_0\mu+\eta_1^2}{\eta_0+\mu}=\dfrac{2\mu(\eta_0^2-\eta_1^2)}
{\eta_0^2-\mu^2},
$$
$$
\psi(\eta_0,\mu)=-F(\eta_0,\mu)+F(-\eta_0,\mu)
=-\dfrac{\eta_0\mu-\eta_1^2}{\eta_0-\mu}+
\dfrac{\eta_0\mu+\eta_1^2}{\eta_0+\mu}=\dfrac{2\eta_0(\eta_1^2-\mu^2)}
{\eta_0^2-\mu^2}.
$$

This solution is naturally called as mode of Debay (this is plasma
mode). In the case of low frequencies it describes well-known
screening of Debay \cite{Abrikosov}. The external field penetrates
into plasma on the depth of $r_D,\; r_D$ is the raduis of Debay.
When the external field frequencies are close to Langmuir
frequencies, the mode of Debay describes plasma oscillations (see,
for instance, \cite{Abrikosov,16}).

\textbf{Note 5.1}.  If to enter expression $c/z $ "inside" of
expression of dispersion function we will receive expression for
dispersion function $h(z)$ from our article \cite {Lesskis}
$$
h(z)=\dfrac{c}{z}\lambda(z)=\dfrac{c}{z}-\dfrac{1}{2}\int\limits_{-1}^{1}
\dfrac{\mu z-\eta_1^2}{\mu-z}d\mu=
\dfrac{c}{z}-z-(z^2-\eta_1^2)
\dfrac{1}{2}\ln\dfrac{z-1}{z+1}.
$$

\begin{center}
\item{}\section{Expansions by eigen functions}
\end{center}

We will seek for the solution of the system of equations (2.5)
and (2.6) with boundary conditions (3.1), (3.6) and (2.7) in the form
of linear combination of discrete eigen solutions of the
characteristic system and integral taken over continuous spectrum of
the system. Let us prove that the following theorem is true.

\textbf{Theorem 6.1}. {\it System of equations (2.5) and (2.6) with
boundary conditions (3.1), (3.6) and (2.7) has a unique solution,
which can be presented as an expansion by eigen functions of the
characteristic system}
$$
h(x,\mu)=\dfrac{E_\infty}{w_0}\mu+\dfrac{E_0}{w_0}\Bigg[
 \dfrac{\eta_0\mu-\eta_1^2}{\eta_0-\mu}
 \exp\big(-\dfrac{w_0x}{\eta_0}\big)+
 \dfrac{\eta_0\mu+\eta_1^2}{\eta_0+\mu}
 \exp\big(\dfrac{w_0x}{\eta_0}\big)\Bigg]+$$$$+
 \intl_{-1}^{1}
 \Bigg[\exp\big(-\dfrac{w_0x}{\eta}\big) F(\eta,\mu)+
 \exp\big(\dfrac{w_0x}{\eta}\big)F(-\eta,\mu)\Bigg]
 \dfrac{E(\eta)}{w_0}\,d\eta,
  \eqno{(6.1)}
 $$
$$
 e(x)=E_\infty+E_0\Big[\exp\big(-\dfrac{w_0x}{\eta}\big)+
 \exp\big(\dfrac{w_0x}{\eta_0}\big)\Big]+
$$
$$
+\intl_{-1}^{1}\Bigg[
 \exp\big(-\dfrac{w_0x}{\eta}\big)+
\exp\big(\dfrac{w_0x}{\eta}\big)\Bigg]E(\eta)\,d\eta.
 \eqno{(6.2)}
 $$

{\it Here $E_0$ and  $E_\infty$ is unknown coefficients corresponding 
to the discrete spectrum ($E_0$ is
the amplitude of Debay, $E_1$ is the amplitude of Drude),
$E(\eta)$ is unknown function, which is called as coefficient
of continuous spectrum}.

When $(\Omega,\varepsilon)\in D^-$ in expansions (6.1) and (6.2) we
should take $E_0=0$. 

Further we will consider the following case $(\Omega,\varepsilon)\in
D^+$.

Our purpose is to find the coefficient of the continuous spectrum,
coefficients of the discrete spectrum and to built expressions for
electron distribution function at the plasma surface and electric
field.

{\bf Proof.}
Let us notice, that the formula (6.1) can be transformed and to such form
$$
h(x,\mu)=\dfrac{E_\infty}{w_0}\mu+\dfrac{2E_0}{w_0}\Bigg[
 \dfrac{\eta_0(\mu^2-\eta_1^2}{\mu^2-\eta_0^2}
 \sh\dfrac{w_0x}{\eta_0}-
 \dfrac{\mu(\eta_0^2-\eta_1^2)}{\mu^2-\eta_0^2}
 \ch\dfrac{w_0x}{\eta_0}\Bigg]+
$$
$$
+\dfrac{1}{w_0}\int\limits_{-1}^{1}\Bigg[\ch\dfrac{w_0x}{\eta}F(\eta,\mu)+
\sh\dfrac{w_0x}{\eta}F(-\eta,\mu)\Bigg]E(\eta)d\eta.
\eqno{(6.1')}
$$

Let us consider expansion (6.1), we will replace in it
$ \mu $ on $-\mu $. Then we will substitute the difference 
$h (1, \mu)-h (1,-\mu) $ in a boundary condition (3.1). 
After variety of transformations let us have
$$
E_\infty\mu+
E_0\Big[F(\eta_0,\mu)+F(-\eta_0,\mu)\Big]\ch\dfrac{w_0}{\eta_0}+
$$
$$
+\int\limits_{-1}^{1}F(\eta,\mu)E(\eta)\ch\dfrac{w_0}{\eta}d\eta=
\dfrac{z_0A_1}{2}\Big(\mu-\dfrac{2}{3}\Big),\quad 0<\mu<1.
\eqno{(6.3)}
$$

Substituting expansion (6.2) in (2.7), we will have
$$
E_\infty+2E_0\ch\dfrac{w_0}{\eta_0}+2\int\limits_{-1}^{1}
E(\eta)\ch\dfrac{w_0}{\eta}\,d\eta=1.
\eqno{(6.4)}
$$

Let us pass from Fredholm integral equation  (6.3) to
singular integral equation with Cauchy kernel, having substituted in
(6.3) obvious representation $F (\eta, \mu) $
$$
E_\infty\mu+E_0\varphi(\eta_0,\mu)\ch\dfrac{w_0}{\eta_0}+
\int\limits_{-1}^{1}
\dfrac{\mu\eta-\eta_1^2}{\eta-\mu}
E(\eta)\ch\dfrac{w_0}{\eta}\,d\eta-2c\dfrac{\lambda(\mu)}{\mu}
\ch\dfrac{w_0}{\mu}\,E(\mu)=
$$
$$
=\dfrac{w_0A_1}{2}\Big(\mu-\dfrac{2}{3}\Big),\qquad 0<\mu<1,
\eqno{(6.5)}
$$
where
$$
\varphi(\eta_0,\mu)=F(\eta_0,\mu)+F(-\eta_0,\mu)=
\dfrac{\eta_0\mu-\eta_1^2}
{\eta_0-\mu}+\dfrac{\eta_0\mu+\eta_1^2}{\eta_0+\mu}
=
2\mu\dfrac{\eta_0^2-\eta_1^2}{\eta_0^2-\mu^2}.
$$

It is easy to check up, that function
$$
M(z)=\int\limits_{-1}^{1}\dfrac{z\eta-\eta_1^2}{\eta-z}
\ch\dfrac{w_0}{\eta}\,E(\eta)d\eta
\eqno{(6.6)}
$$
is odd. Besides, all members of the equation (6.5) are
odd on $ \mu $, except one member from its right part
$-\dfrac{1}{3}w_0A_1$.

Hence, the equation (6.5) can be extended in the interval
$-1 <\mu <1$ in the next symmetric form
$$
E_\infty\mu+E_0\varphi(\eta_0,\mu)\ch\dfrac{w_0}{\mu}+
\int\limits_{-1}^{1}
\ch\dfrac{w_0}{\eta}\,\dfrac{\mu\eta-\eta_1^2}{\eta-\mu}
E(\eta)d\eta-$$$$-2c\dfrac{\lambda(\mu)}{\mu}
\ch\dfrac{w_0}{\mu}\,E(\mu) -\dfrac{w_0A_1}{2}\mu=
-\dfrac{1}{3}w_0A_1\sign \mu,\quad -1<\mu<1.
\eqno{(6.7)}
$$\medskip

Let us reduce the equation (6.7) to Riemann --- Hilbert boundary
value problem. For this purpose we will take advantage Sohkotsky
formulas for the auxiliary functions $M(z)$ and dispersion function 
$\lambda(z)$
$$
M^+(\mu)-M^-(\mu)=2\pi i\ch\dfrac{w_0}{\mu}\,(\mu^2-\eta_1^2)
E(\mu), \quad -1<\mu<1,
\eqno{(6.8)}
$$
$$
\dfrac{M^+(\mu)+M^-(\mu)}{2}=M(\mu),\qquad -1<\mu<1,
$$
where
$$
M(\mu)=\int\limits_{-1}^{1}\dfrac{\mu\eta-\eta_1^2}{\eta-\mu}
\ch\dfrac{w_0}{\eta}\,E(\eta)d\eta,
$$
besides, last integral is understood as singular in the sense of principal
value of Cauchy, and
$$
\lambda^+(\mu)-\lambda^-(\mu)=-\dfrac{i\pi}{c}\mu(\mu^2-\eta_1^2),\quad
\dfrac{\lambda^+(\mu)+\lambda^-(\mu)}{2}=\lambda(\mu),\qquad-1<\mu<1,
$$
where
$$
\lambda(\mu)=1-\dfrac{\mu}{2c}\int\limits_{-1}^{1}\dfrac{\mu'\mu-\eta_1^2}
{\mu'-\mu}d\mu',
$$
besides, last integral is understood as singular in the sense of principal
value of Cauchy also.

As result of use of last formulas we will come to the boundary value 
problem
$$
E_\infty\mu+E_0\varphi(\eta_0,\mu)\ch \dfrac{w_0}{\eta_0}+
\dfrac{1}{2}\big[M^+(\mu)+M^-(\mu)\big]+
\dfrac{1}{2}\dfrac{\lambda^+(\mu)+\lambda^-(\mu)}{\lambda^+(\mu)-
\lambda^-(\mu)}\big[M^+(\mu)-M^-(\mu)\big]-
$$
$$
-\dfrac{w_0A_1}{2}\mu=
-\dfrac{1}{3}w_0A_1\sign \mu,\quad -1<\mu<1.
$$

We transform this equation to the form
$$
\dfrac{1}{2}(M^++M^-)(\lambda^+-\lambda^-)+\dfrac{1}{2}(M^+-M^-)
(\lambda^++\lambda^-)+
$$
$$
+(\lambda^+-\lambda^-)\Big(E_\infty\mu+
E_0\varphi(\eta_0,\mu)\ch \dfrac{w_0}{\eta_0}-
\dfrac{w_0A_1}{2}\mu\Big)=
$$
$$
=\dfrac{1}{3}w_0A_1\sign \mu\dfrac{i\pi}{c}\mu
(\mu^2-\eta_1^2),\qquad -1<\mu<1.
$$

From here we receive the following boundary condition of boundary value
Riemann --- Hilbert
$$
\lambda^+(\mu)\Big[M^+(\mu)+E_\infty\mu+E_0\varphi(\eta_0,\mu)
\ch \dfrac{w_0}{\eta_0}-
\dfrac{w_0A_1}{2}\mu\Big]-
$$
$$
=\lambda^-(\mu)\Big[M^-(\mu)+E_\infty\mu+E_0\varphi(\eta_0,\mu)
\ch \dfrac{w_0}{\eta_0}-
\dfrac{w_0A_1}{2}\mu\Big]=
$$$$
=\dfrac{\pi i}{3c}w_0A_1 \mu\sign \mu (\mu^2-\eta_1^2), \qquad
-1<\mu<1.
\eqno{(6.9)}
$$

We rewrite this problem in the form
$$
\Phi^+(\mu)-\Phi^-(\mu)=\dfrac{i\pi }{3c}
w_0A_1\mu(\mu^2-\eta_1^2)\sign \mu, \quad -1<\mu<1.
\eqno{(6.10)}
$$

In the problem (6.10) $\Phi^{\pm}(\mu)$ are boundary values on
interval  $-1<\mu<1$ of function
$$
\Phi(z)= \lambda^+(z)\Big[M^+(z)+E_\infty z+E_0\varphi(\eta_0,z)
\ch \dfrac{w_0}{\eta_0}-\dfrac{w_0A_1}{2}z\Big],
$$
which is analytic in complex plane with cut $\mathbb{C}\setminus [-1,1]$.

The problem (6.10) is a problem of special case about jump. The jump problem
is the problem of finding of analytical function by its jump on the contour
$L$:
$$
\Phi^+(\mu)-\Phi^+(\mu)=\varphi(\mu), \qquad \mu\in L.
$$

The solution of such problems in a class of functions 
decreasing at infinitely remote point it is given by integral of
Caucy type
$$
\Phi(z)=\dfrac{1}{2\pi i}\int\limits_{L}\dfrac{\varphi(\tau)d\tau}
{\tau-z}.
$$

However, ih the problem (6.10), when
$$
\varphi(\mu)=\dfrac{2i\pi }{6c}
w_0A_1\mu(\mu^2-\eta_1^2)\sign \mu, \quad -1<\mu<1,
$$
unknown function $\Phi(z)$ has at infinitely remote point the following
asymptotic
$$
\Phi(z)=O(z), \qquad z\to \infty.
$$

Therefore it is necessary to search the solution of the problem (6.11)  
in a class of the growing as $z $ in the vicinity of infinitely remote point.

According to \cite{17} the general solution of problem (6.10) 
it is given by the formula
$$
\Phi(z)=\dfrac{1}{2\pi i}\int\limits_{L}\dfrac{\varphi(\tau)d\tau}
{\tau-z}+C_1z.
$$

In explicit form the general solution of problem (6.10) write down
the following form
$$
\lambda(z)\Big[M(z)+E_0\varphi(\eta_0,z)\ch \dfrac{w_0}{\eta_0}+
E_\infty w_0-\dfrac{w_0A_1}{2}z\Big]
=$$$$+\dfrac{1}{3}(w_0A_1)\dfrac{1}{2c}\int\limits_{-1}^{1}
\dfrac{\mu(\mu^2-\eta_1^2)\sign \mu}{\mu-z}d\mu+C_1z,
$$
where $C_1$ is the arbitrary constant.

Let us enter auxiliary function
$$
T(z)=\dfrac{1}{2c}\int\limits_{-1}^{1}
\dfrac{\mu(\mu^2-\eta_1^2)\sign \mu}{\mu-z}d\mu.
$$

Then the general solution  receives the following form
$$
\lambda(z)\Big[M(z)+E_0\varphi(z)+E_\infty z-\dfrac{w_0A_1}{2}z\Big]=
\dfrac{1}{3}w_0A_1T(z)+C_1z.
$$

From this general solution we can find the function $M(z)$
$$
M(z)=-E_\infty z-E_0\varphi(\eta_0,z)\ch \dfrac{w_0}{\eta_0}+
\dfrac{w_0A_1}{2}z+
\dfrac{1}{3}w_0A_1\dfrac{T(z)}{\lambda(z)}+\dfrac{C_1z}{\lambda(z)}.
\eqno{(6.11)}
$$

Let us remove a pole at the solution (6.11) at infinitely remote point.
Let us notice, that the function
$$
\varphi(\eta_0,z)=\dfrac{2z(\eta_1^2-\eta_0^2)}
{z^2-\eta_0^2}.
$$
at $z\to \infty$ has the following asymptotic
$$
\varphi(\eta_0,z)=O\Big(\dfrac{1}{z}\Big), \qquad z\to \infty.
$$
Considering, that function $T(z)$ has exactly the same asymptotic at
$$
C_1=(E_\infty-\dfrac{z_0A_1}{2})\lambda_\infty.
\eqno{(6.12)}
$$

\begin{center}
\item{}\section{Coefficients of discrete and continuous spectra}
\end{center}

Now we will remove poles at the solution (6.11) at the points $ \pm \eta_0$.
Let us allocate in the right part of the solution (6.11) members 
containing the polar singularity at the point $z =\eta_0$. 
In the point vicinity $z =\eta_0$
taking into account equality $\lambda(\eta_0)=0$ it is carried out 
the following  expansion
$$
M(z)=-\Big(E_\infty-\dfrac{w_0A_1}{2}\Big)z-
E_0\dfrac{\eta_1^2+\eta_0z}{z+\eta_0}\ch\dfrac{w_0}{\eta_0}+
$$
$$
+\dfrac{1}{z-\eta_0}\Bigg[-E_0(\eta_1^2-\eta_0^2)\ch\dfrac{w_0}{\eta_0}+
\dfrac{(1/3)w_0A_1T(z)+\eta_0(E_\infty-w_0A_1/2)\lambda_\infty}
{\lambda'(\eta_0)+(1/2!)\lambda''(\eta_0)(z-\eta_0)+\cdots}\Bigg].
$$

From here it is visible, that for pole elimination at the point $z =\eta_0$
it is necessary to equate to zero expression in a square bracket,
calculated at $z =\eta_0$. Then we receive, that
$$
w_0A_1=
\dfrac{E_0\lambda'(\eta_0)(\eta_1^2-\eta_0^2)\ch(w_0/\eta_0)-\eta_0
E_\infty\lambda_\infty}
{T(\eta_0)/3-\eta_0\lambda_\infty/2},
\eqno{(7.1)}
$$
therefore
$$
2E_0\ch \dfrac{w_0}{\eta_0}=\dfrac{w_0A_1[2T(\eta_0)/3-
\lambda_\infty\eta_0]+
2\eta_0\lambda_\infty E_\infty}
{\lambda'(\eta_0)(\eta_1^2-\eta_0^2)}.
\eqno{(7.2)}
$$

The continuous spectra coefficient $E(\eta)$ we find from formulas
(6.11) and (6.12)
$$
E(\eta)=\dfrac{1}{2\pi i}\cdot \dfrac{M^+(\eta)-M^-(\eta)}
{\ch\dfrac{w_0}{\eta}\,(\eta^2-\eta_1^2)}.
\eqno{(7.3)}
$$

Difference $M^+(\eta)-M^-(\eta)$ from the formula (7.3) we will find 
with the help formulas of the general solution (6.12). As result 
we receive the following expression
$$
E(\eta)=\dfrac{1}{2\pi i \ch\dfrac{w_0}{\eta}\,(\eta^2-\eta_1^2)}
\Bigg\{\dfrac{1}{3}w_0A_1\Bigg[\dfrac{T^+(\eta)}{\lambda^+(\eta)}-
\dfrac{T^-(\eta)}{\lambda^-(\eta)}\Bigg]+
$$
$$
+\big(E_\infty-\dfrac{w_0A_1}{2}\big)\lambda_\infty \eta
\Bigg[\dfrac{1}{\lambda^+(\eta)}-
\dfrac{1}{\lambda^-(\eta)}\Bigg]\Bigg\}.
\eqno{(7.4)}
$$

Let us substitute equalities (7.2) and (7.4) in the equation on the 
field (6.4).  We receive the following equation
$$
\lambda_\infty E_\infty\Big(\dfrac{1}{\lambda_\infty}-
\dfrac{2\eta_0}{\lambda'(\eta_0)(\eta_0^2-\eta_1^2)}+J_1\Big)+
w_0A_1\Big(-\dfrac{2T(\eta_0)/3-\lambda_\infty\eta_0}
{\lambda'(\eta_0)(\eta_0^2-\eta_1^2)}-\dfrac{\lambda_\infty}{2}J_1+
\dfrac{1}{3}J_2\Big)=1,
\eqno{(7.5)}
$$
where
$$
J_1=\dfrac{1}{2\pi i}\int\limits_{-1}^{1}\Bigg(\dfrac{1}{\lambda^+(\eta)}-
\dfrac{1}{\lambda^-(\eta)}\Bigg)\dfrac{\eta\,d\eta}{\eta^2-\eta_1^2},
$$
$$
J_2=\dfrac{1}{2\pi i}\int\limits_{-1}^{1}
\Bigg(\dfrac{T^+(\eta)}{\lambda^+(\eta)}-
\dfrac{T^-(\eta)}{\lambda^-(\eta)}\Bigg)
\dfrac{d\eta}{\eta^2-\eta_1^2}.
$$
Integrals $J_1$ and $J_2$ from the equation (7.5) can be calculated
analytically by means of the theory of residues and contour integration.
For the first integral we have
$$
J_1=\dfrac{1}{2\pi i}\int\limits_{-1}^{1}\Big[\dfrac{1}{\lambda^+(\eta)}-
\dfrac{1}{\lambda^-(\eta)}\Big]\dfrac{\eta
d\eta}{\eta^2-\eta_1^2}=
$$
$$
=\Big[\Res_{\infty}+\Res_{\eta_1}+\Res_{-\eta_1}+\Res_{\eta_0}+
\Res_{-\eta_0}\Big]\dfrac{z}{(z^2-\eta_1^2)\lambda(z)}.
$$

Let us notice, that
$$
\Res_{\infty}\dfrac{z}{(z^2-\eta_1^2)\lambda(z)}=
-\dfrac{1}{\lambda_\infty},\qquad
\Res_{\pm\eta_1}\dfrac{z}{(z^2-\eta_1^2)\lambda(z)}=
\dfrac{1}{2\lambda(\eta_1)},
$$
$$
\Res_{\pm \eta_0}\dfrac{z}{(z^2-\eta_1^2)\lambda(z)}=\dfrac{\eta_0}
{(\eta_0^2-\eta_1^2)\lambda'(\eta_0)}.
$$

Hence, the integral is equal
$$
J_1=-\dfrac{1}{\lambda_\infty}+\dfrac{1}{\lambda_1}+\dfrac{2\eta_0}
{(\eta_0^2-\eta_1^2)\lambda'(\eta_0)},
$$
where
$$
\lambda_1=\lambda(\eta_1)=1-\dfrac{1}{z_0}.
$$

In the same way we calculate the second integral
$$
J_2=\dfrac{1}{2\pi i}\int\limits_{-1}^{1}
\dfrac{T_2(\eta)}{\eta^2-\eta_1^2}\Big[\dfrac{1}{\lambda^+(\eta)}-
\dfrac{1}{\lambda^-(\eta)}\Big]d\eta=
$$
$$
=\Big[\Res_{\infty}+\Res_{\eta_1}+\Res_{-\eta_1}+\Res_{\eta_0}+
\Res_{-\eta_0}\Big]\dfrac{T(z)}{(z^2-\eta_1^2)\lambda(z)}.
$$

Let us notice, that
$$
\Res_{\infty}\dfrac{T(z)}{(z^2-\eta_1^2)\lambda(z)}=0,\qquad
\Res_{\pm\eta_1}\dfrac{T(z)}{(z^2-\eta_1^2)\lambda(z)}=\pm
\dfrac{T(\pm\eta_1)}{2\eta_1\lambda_1},
$$
$$
\Res_{\pm\eta_0}\dfrac{T(z)}{(z^2-\eta_1^2)\lambda(z)}=
\pm \dfrac{T(\pm\eta_0)}{(\eta_0^2-\eta_1^2)\lambda'(\eta_0)}.
$$

Hence, the integral is equal
$$
J_2=\dfrac{T(\eta_1)-T(-\eta_1)}{2\eta_1\lambda_1}+
\dfrac{T(\eta_0)-T(-\eta_0)}{(\eta_0^2-\eta_1^2)\lambda'(\eta_0)}.
$$
Let us notice, that
$
T(\eta_1)-T(-\eta_1)=
\dfrac{\eta_1}{c}.
$
Thus, it is definitively received
$$
J_2=\dfrac{1}{2c\lambda_1}+
\dfrac{2T(\eta_0)}{(\eta_0^2-\eta_1^2)\lambda'(\eta_0)}.
$$

By means of these integrals the condition on the field (7.5) 
transforms to the form
$$
\dfrac{\lambda_\infty E_\infty}{\lambda_1}+\dfrac{w_0A_1}{2\lambda_1}
\Big(\dfrac{1}{3c}+\lambda_1-\lambda_\infty\Big)=1,
\quad\text{or}\quad \dfrac{\lambda_\infty E_\infty}{\lambda_1}=1,
$$
because $\lambda_\infty=\lambda_1+1/3c$.

Hence, from last equation we find the  Drude amplitude 
$$
E_\infty=\dfrac{\lambda_1}{\lambda_\infty}.
\eqno{(7.6)}
$$

Now the condition (7.2) will be transformed to the form
$$
E_0=-\dfrac{\lambda_1\eta_0+w_0A_1(\frac{1}{3}T(\eta_0)-
\frac{1}{2}\lambda_\infty\eta_0)}
{\lambda'(\eta_0)(\eta_0^2-\eta_1^2)\ch({w_0}/{\eta_0})}.
\eqno{(7.7)}
$$

\begin{center}
\item{}\section{Integral condition on distribution function} 
\end{center}

Let us notice, that at transition through the positive part of the cut
$(0,1)$ functions
$T(z)$ and $\lambda(z)$ make the jump, differing only by signs.
Really, the formula for $T(z)$ we will present in the kind
$$
T(z)=\dfrac{1}{2c}\int\limits_{0}^{1}\mu
(\mu^2-\eta_1^2)\Bigg[\dfrac{1}{\mu-z}-\dfrac{1}{\mu+z}\Bigg]d\mu=
\dfrac{z}{c}\int\limits_{0}^{1}\dfrac{\mu(\mu^2-\eta_1^2)}
{\mu^2-z^2}d\mu,
$$
or
$$
T(z)=\dfrac{z}{2c}\int\limits_{0}^{1}
(\mu^2-\eta_1^2)\Bigg[\dfrac{1}{\mu-z}+\dfrac{1}{\mu+z}\Bigg]d\mu=
\dfrac{z}{c}\int\limits_{0}^{1}\dfrac{\mu(\mu^2-\eta_1^2)}
{\mu^2-z^2}d\mu.
$$

This integral is easy for calculating in an explicit form.
In a complex plane with a cut function $T(z)$ is calculated by
the formula
$$
T(z)=\dfrac{z}{2c}\Big[1+(z^2-\eta_1^2)
\ln\Big(1-\dfrac{1}{z^2}\Big)\Big], \qquad z\in \mathbb{C},
$$
and on the cut this integral is calculated by formula
$$
T(\eta)=\dfrac{\eta}{2c}\Big[1+(\eta^2-\eta_1^2)
\ln\Big(\dfrac{1}{\eta^2}-1\Big)\Big], \qquad -1<\eta<+1.
$$

Now from the Sohkotsky formula for a difference of boundary values
we receive, that at  $0<\eta<1$
$$
\lambda^+(\eta)=\lambda(\eta)\pm \dfrac{i\pi }{2c}\eta(\eta_1^2-\eta^2),
\qquad
T^+(\eta)=T(\eta)\pm \dfrac{i\pi}{2c} \eta
(\eta^2-\eta_1^2)\sign \eta.
$$

Now it is easy to find, that
$$
T^+(\eta)\lambda^-(\eta)-T^-(\eta)\lambda^+(\eta)=
\big[T(\eta)+\lambda(\eta)\sign \eta\big]\cdot \dfrac{i\pi }{c}\eta
(\eta^2-\eta_1^2).
$$

We enter the integral
$$
T_0(z)= \dfrac{1}{2c}\int\limits_{0}^{1}
\dfrac{\eta^2-\eta_1^2}{\eta-z}d\eta.
$$
and we denote $T_1(\eta)=T(\eta)+\lambda(\eta)\sign \eta$. It is clear 
that $T_1(\eta)$ is odd function, besides
$$
T_1(\eta)=
\Bigg\{
\begin{array}{c}
         1+2\eta T_0(-\eta),\quad \eta>0, \\
         -(1-2\eta T_0(\eta)), \quad \eta<0. 
       \end{array}
$$

Substituting last two equalities in (7.3), we receive the formula for
calculations of coefficient of continuous spectrum
$$
E(\eta)=\dfrac{1}{2c\ch\big(w_0/\eta\big)}
\Bigg[\dfrac{\lambda_1\eta^2}{\lambda^+(\eta)\lambda^-(\eta)}+
\dfrac{z_0A_1}{6}\dfrac{2\eta T_1(\eta)-3\lambda_\infty \eta^2}
{\lambda^+(\eta)\lambda^-(\eta)}\Bigg].
\eqno{(8.1)}
$$

Clearly that in complex plane the integral $T_0(z)$  is calculated under 
the formula
$$
T_0(z)=\dfrac{1}{2c}\Big[\dfrac{1}{2}+z+(z^2-\eta_1^2)
\ln\Big(1-\dfrac{1}{z}\Big)\Big],
$$
and at $0<\eta<1$ it is calculated by formula
$$
T_0(\eta)=\dfrac{1}{2c}\Big[\dfrac{1}{2}+\eta+(\eta^2-\eta_1^2)
\ln\Big(\dfrac{1}{\eta}-1\Big)\Big].
$$

By means of this function we will present dispersion function
in the form
$$
\lambda(z)=1-zT_0(z)+zT_0(-z),
$$
the function $T(z)$ we express by this integral also
$$
T(z)=zT_0(z)+zT_0(-z).
$$
The sum of two last expressions is equal
$$
\lambda(z)+T(z)=1+2zT_0(-z).
$$

We will notice, that the integral $T(-z)$ is not singul on the cut
$0 <\eta <1$. The sum $\lambda(\eta) + T(\eta) \sign\eta$ on the cut 
of $0 <\eta <1$ it is calculated in an explicit form without quadratures
$$
\lambda(\eta)+T(\eta)\sign \eta=1+2\eta T_0(-\eta)=
$$
$$
=1+\dfrac{1}{2c}\Bigg[\eta-
2\eta^2+2\eta(\eta^2-\eta_1^2)\ln\Big(\dfrac{1}{\eta}+1\Big)\Bigg].
$$

Now we consider integral boundary condition (3.6).
We rewrite it in the form
$$
\int\limits_{0}^{1}(\mu^2-\dfrac{2}{3}\mu)h(-a,\mu)d\mu=
-\dfrac{1-\alpha_p}{\alpha_p}\cdot \dfrac{A_1}{36}.
\eqno{(8.2)}
$$

Here, as it is easy to see, 
$$
h(-a,\mu)=\dfrac{E_\infty}{w_0}\mu+\dfrac{E_0}{w_0}\Big[F(\eta_0,\mu)
e^{w_0/\eta_0}+F(-\eta_0,\mu)e^{-w_0/\eta_0}\Big]+$$$$+
2\int\limits_{-1}^{1}F(\eta,\mu)E(\eta)e^{w_0/\eta}d\eta.
\eqno{(8.3)}
$$

Let us substitute (7.4) in expansion (7.5) for the function 
$h(x,\mu)$. We receive the following equation
$$
\dfrac{E_\infty}{36}+E_0m(\eta_0)+
2\int\limits_{-1}^{1}e^{w_0/\eta}m(\eta)E(\eta)d\eta=
-\dfrac{1-\alpha_p}{\alpha_p}\cdot \dfrac{w_0A_1}{36},
\eqno{(8.4)}
$$
where
$$
m(\eta_0)=e^{w_0/\eta_0}m_0(\eta_0)+e^{-w_0/\eta_0}m_0(-\eta_0).
$$

Besides, in the equation (8.4) following designations are accepted
$$
m_0(\pm \eta_0)=\int\limits_{0}^{1}\Big(\mu^2-\dfrac{2}{3}\mu\Big)
F(\pm \eta_0,\mu)d\mu,\qquad
m(\eta)=\int\limits_{0}^{1}\Big(\mu^2-\dfrac{2}{3}\mu\Big)
F(\eta,\mu)d\mu.
$$

Let us substitute in the equation (8.4) the Debaye amplitude $E_0$, 
defined by equation (7.7), and continuous spectrum  coefficient 
$E(\eta)$, defined by equality (8.1). We receive the equation
$$
\lambda_1\Big(\dfrac{1}{36\lambda_\infty}-\eta_0\beta_0+I_0\Big)+
w_0A_1\Big[-\Big(\dfrac{1}{3}T(\eta_0)-\dfrac{\lambda_\infty}{2}\Big)
\beta_0+\dfrac{1}{3}I_1-\dfrac{\lambda_\infty}{2}I_0+
\dfrac{1-\alpha_p}{36\alpha_p}\Big]=0.
\eqno{(8.5)}
$$

In the equation (8.5) are entered the following  designations
$$
\beta_0=\dfrac{m_0(\eta_0)e^{w_0/\eta_0}+m(-\eta_0)e^{-w_0/\eta_0}}
{\lambda'(\eta_0)(\eta_0^2-\eta_1^2)\ch(w_0/\eta_0)},
$$
$$
\beta(\eta)=\dfrac{m(\eta)e^{w_0/\eta}}{\lambda^+(\eta)\lambda^-(\eta)
\ch (w_0/\eta)},\qquad
I_0=\dfrac{1}{c}\int\limits_{-1}^{1}\eta^2\beta(\eta)d\eta, \qquad
I_1=\dfrac{1}{c}\int\limits_{-1}^{1}\eta T_1(\eta)\beta(\eta)d\eta.
$$

Now from equation (8.5) we find the constant $A_1$
$$
z_0A_1=-\dfrac{\lambda_1\Big(\frac{1}{36\lambda_\infty}-
\eta_0\beta_0+I_0\Big)}{\frac{1}{3}I_1-\frac{\lambda_\infty}{2}I_0+
\frac{1-\alpha_p}{36\alpha_p}-\frac{\beta_0}{3}T(\eta_0)+
\frac{\beta_0}{2}\lambda_\infty \eta_0}.
\eqno{(8.6)}
$$

The solving of this initial boundary problem on it is finished. 

Let us calculate integrals $m_0(\pm\eta_0)$ and $m(\eta)$ in an explicit 
form. Integrals $m_0 (\eta_0) $ and $m_0(-\eta_0)$ are easily calculated
$$
m_0(\eta_0)=(\eta_1^2-\eta_0^2)\Bigg[\Big(-\dfrac{1}{6}+\eta_0\Big)+
\Big(\eta_0^2-\dfrac{2}{3}\eta_0\Big)\ln
\Big(1-\dfrac{1}{\eta_0}\Big)\Bigg],
$$
$$
m_0(-\eta_0)=(\eta_1^2-\eta_0^2)\Bigg[\Big(-\dfrac{1}{6}-\eta_0\Big)+
\Big(\eta_0^2+\dfrac{2}{3}\eta_0\Big)\ln
\Big(1+\dfrac{1}{\eta_0}\Big)\Bigg],
$$

We calculate the integral $m(\eta)$
$$
m(\eta)=\int\limits_{0}^{1}\Big(\mu^2-\dfrac{2}{3}\mu\Big)
\Big[P\dfrac{\mu\eta-\eta_1^2}{\eta-\mu}-2c\dfrac{\lambda(\eta)}{\eta}
\delta(\eta-\mu)\Big]d\mu=
$$
$$
=\int\limits_{0}^{1}\dfrac{(\mu^2-\frac{2}{3}\mu)(\eta_1^2-\mu\eta)}
{\mu-\eta}d\mu-2c(\eta-\dfrac{2}{3})\lambda(\eta)\theta_+(\eta).
$$
Here $\theta_+(\eta)$ is the characteristical function of interval
$(0,1)$, i.e.
$$
\theta_+(\eta)=\left\{\begin{array}{c}
                        1,\quad 0<\eta<1, \\
                        0,\quad -1<\eta<0. 
                      \end{array}
\right\}
$$

From here we receive, that at $0<v<1$
$$
m(\eta)=(\eta^2-\eta_1^2)(\dfrac{1}{6}-\eta)-(\eta^2-\eta_1^2)(\eta^2-
\dfrac{2}{3}\eta)\ln(1+\dfrac{1}{\eta})+
2(\eta^2-c)(\eta-\dfrac{2}{3}),
$$
and at $-1<\eta<0$
$$
m(\eta)=(\eta^2-\eta_1^2)\Big[\dfrac{1}{6}-\eta-(\eta^2-\dfrac{2}{3}\eta)
\ln(1-\dfrac{1}{\eta})\Big].
$$

Last two formulas it is possible to unite in one
$$
m(\eta)=(\eta^2-\eta_1^2)\Big[\dfrac{1}{6}-\eta-(\eta^2-\dfrac{2}{3}\eta)
f_+(\eta)\Big]+2(\eta^2-c)(\eta-\dfrac{2}{3})\theta_+(\eta),
$$
where
$$
f_+(\eta)=\left\{\begin{array}{c}
                   \ln\dfrac{1+\eta}{\eta},\quad 0<\eta<1, \\
                   \ln\dfrac{1-\eta}{\eta},\quad -1<\eta<0. 
                 \end{array}
\right\}
$$

\begin{center}
\item \section{Energy absorption in layer}
\end{center}

The energy of an electromagnetic wave absorbed in the slab of degenerate
plasma bi well-known formula \cite{LandauFK}
$$
Q=\dfrac{1}{2}\Re \int\limits_{-a}^{a}j(x)E^*(x)\,dx.
\eqno{(9.1)}
$$
Here "asterisk" \, means complex conjugation, and
$j(x)$ is the density of a current,
$$
j(x)=\dfrac{1}{2}\int\limits_{-1}^{1}\mu h(x,\mu)\,d\mu.
$$

In view of one-dimensionality of a problem the equation for electric 
field has the form  ${dE}/{dx} =4\pi q $. All quantities have 
dependence from time $\exp (-i\omega t)$, i.e. $E=E(x)\exp(-i\omega t)$, 
and etc. The  continuity equation for system a charge -- current 
will be rewrite so: ${dj(x)}/{dx}-i\omega q(x) =0$. From last two
equations we have
$$
i\omega \dfrac{dE}{dx}=4\pi \dfrac{dj(x)}{dx}.
$$
Integrating it equality and considering equality to current zero on border, 
we have
$$
j(x)=\dfrac{i\omega}{4\pi}\Big(E(x)-E_0\Big),
\eqno{(9.2)}
$$
where $E_0$ is the amplitude  of an external field on border
(real quantity).

Subsituting (9.2) into (9.1), we find
$$
Q=\dfrac{\omega}{8\pi}\Re\int\limits_{-a}^{a}i
(E(x)-E_0)E^*(x)\,dx.
\eqno{(9.3)}
$$

We notice, that $EE^*=|E|^2>0$ is the real quantity, therefore,
$$
\Re\{i(E-E_0)E^*\}=-E_0\Re(iE^*)=$$$$=-E_0\Re(i(\Re E-
i \Im E))=-E_0\Im E(x).
$$
Hence, according to (9.3) we have
$$
Q=-\dfrac{\omega E_0}{8\pi}\int\limits_{-a}^{a}\Im E(x)\,dx,
$$
or, by dimensionless electric field,
$$
Q=-\dfrac{\omega a
E_0^2}{8\pi}\Im\int\limits_{-1}^{1}e(x_1)\,dx_1.
\eqno{(9.4)}
$$

Here the expression received above for electric field is used, 
which we will present in the form
$$
e(x_1)=\dfrac{\lambda_1}{\lambda_\infty}+2E_0\ch\dfrac{w_0x_1}{\eta_0}+
2\int\limits_{-1}^{1}\ch\dfrac{w_0x_1}{\eta}E(\eta)d\eta.
\eqno{(9.5)}
$$
Coefficients of continuous and discrete spectra of expansion (9.5) 
have been calculated in the explicit form.

Let us substitute in the formula (9.4) expression for the electric
field (9.5). After integration it is received, that
$$
Q=\dfrac{\omega a_0 E_0^2}{4\pi}Q_0,
$$
where $Q_0$ is the dimensionless part of absorption, defined by formula
$$
Q_0=-\Im  Q_1,
\eqno{(9.5')}
$$
where
$$
Q_1=\dfrac{\lambda_1}{\lambda_\infty}+
2E_0\dfrac{\eta_0}{w_0}\sh\dfrac{w_0}{\eta_0}
+\dfrac{2}{w_0}
\int\limits_{-1}^{1}\eta E(\eta)\sh \dfrac{w_0}{\eta}d\eta.
\eqno{(9.6)}
$$

The formula (9.5) represents quantity of absorption of 
electric field  energy in the metal layer. It is defined by frequency and
amplitude of an external field $ \omega $ and $E_0$, width of the 
layer $a_0$, and by parametres $w_0, z_0, \varepsilon, \Omega $, 
reflecting properties metal.

We consider integral from (9.6)
$$
J\equiv\int\limits_{-1}^{1}\eta E(\eta)\sh\dfrac{w_0}{\eta}\,d\eta=
\dfrac{\lambda_1-w_0A_1/2}{2c}\int\limits_{-1}^{1}
\dfrac{\eta^3\th(w_0/\eta)\,d\eta}{\lambda^+(\eta)\lambda^-(\eta)}+
\dfrac{w_0A_1}{6c}\int\limits_{-1}^{1}\dfrac{\eta^2\th({w_0}/{\eta})
T_1(\eta)\,d\eta}{\lambda^+(\eta)\lambda^-(\eta)}.
$$

Then taking into account equality
$$
\dfrac{1}{\lambda^+(\eta)\lambda^-(\eta)}=
\dfrac{c}{i\pi\eta(\eta^2-\eta_1^2)}\Big[\dfrac{1}{\lambda^+(\eta)}-
\dfrac{1}{\lambda^-(\eta)}\Big]
$$
we receive
$$
J=\big(\lambda_1-\dfrac{w_0A_1}{2}\big)J_1+\dfrac{w_0A_1}{3}J_0,
$$
where
$$
J_{1}=\dfrac{1}{2\pi i}\int\limits_{-1}^{1}
\dfrac{\eta^2\sh({w_0}/{\eta})}{(\eta^2-\eta_1^2)\ch({w_0}/{\eta})}
\Bigg[\dfrac{1}{\lambda^+(\eta)}-\dfrac{1}{\lambda^-(\eta)}\Bigg]\,d\eta,
$$
$$
J_{0}=\dfrac{1}{2c}\int\limits_{-1}^{1}\dfrac{\eta^2T_1(\eta)
\th({w_0}/{\eta})}{\lambda^+(\eta)\lambda^-(\eta)}
\,d\eta=\dfrac{1}{c}\int\limits_{0}^{1}\dfrac{\eta^2T_1(\eta)
\th({w_0}/{\eta})}{\lambda^+(\eta)\lambda^-(\eta)}\,d\eta.
\eqno{(9.7)}
$$

Let us notice, that integral $J_1$ it is possible to calculate  
analytically, and integral
$J_0$ analytically to calculate it is impossible, since function 
continuation
$T_1(z)$ from the real axis in a complex plane it is carried out
by non--analytical function $ |z | (1+2zT_0 (-|z |))/z $.

Function zero $\ch(w_0/z)$ will be necessary for us
$$
t_k=\dfrac{2w_0i}{\pi(2k+1)},\qquad k=0,\pm 1,\pm 2, \dots.
$$

The integral $J_{1}$ has been calculated earlier in our work
\cite {11}. By means of contour integration and the theory of residues 
it is found, that
$$
J_{1}=\Big[\Res_{\infty}+\Res_{\eta_0}+\Res_{-\eta_0}+\Res_{\eta_1}+
\Res_{-\eta_1}+\sum\limits_{k=-\infty}^{+\infty}\Res_{t_k}\Big]\Phi(z),
$$
where
$$
\Phi(z)=\dfrac{z^2\th(w_0/z)}{\lambda(z)(z^2-\eta_1^2)}.
$$

Calculating all residues from the previous equality, we receive
$$
J_{1}=-\dfrac{w_0}{\lambda_\infty}+\dfrac{2\eta_0^2\th(w_0/\eta_0)}
{\lambda'(\eta_0)(\eta_0^2-\eta_1^2)}+
\dfrac{\eta_1}{\lambda_1}\th\dfrac{w_0}{\eta_1}-\dfrac{1}{w_0}
\sum\limits_{k=-\infty}^{+\infty}
\dfrac{t_k^4}{\lambda(t_k)(t_k^2-\eta_1^2)}.
\eqno{(9.8)}
$$

By means of this expression the formula (9.6) can be presented
in the explicit form
$$
Q_1=\dfrac{\lambda_1}{\lambda_\infty}+2E_0\dfrac{\eta_0}{w_0}
\sh\dfrac{w_0}{\eta_0}+\Big(\dfrac{\lambda_1}{w_0}-\lambda_\infty
\dfrac{A_1}{2}\Big)J_1+\dfrac{A_1}{3}J_0.
\eqno{(9.9)}
$$

In the formula (9.9) Debaye amplitude $E_0$ is calculated according 
to (7.7), quantity $A_1$ is calculated according to (8.6), 
and integrals $J_0$ and $J_1$ are calculated under formulas
(9.7) and (9.8) accordingly.

\begin{center}
\item \section{Conclusions}
\end{center}

In the present paper the linearized problem of plasma oscillations
in layer (particularly, thin films) in external longitudinal 
alternating electric field is solved
analytically. Specular -- accomodative boundary conditions of
electron reflection from the plasma boundaries are considered.
Coefficients of continuous and discrete spectra of the problem are
found, and electron distribution function on the plasma boundary and
electric field are expressed in explicit form.

Separation of variables
leads to characteristic system of the equations. Its solution in
space of the generalized functions allows to find the eigen
solutions of initial system of equations of Boltzmann --- Vlasov ---
Maxwell, correspond to  continuous spectrum.

Then the discrete spectrum of this problem consisting of zero of the 
dispersion equations is investigated. Such zeros are three. 
One zero is infinite remote point of complex plane. 
It is correspond to the eigen solution "Drude mode". 
Others two points of discrete spectrum are differing 
with signs two zero of dispersion function. 
These zero are correspond to the eigen solution named "Debay mode".

It is found out, that on  plane of parametres of a problem
($\Omega, \varepsilon $), where 
$ \Omega =\omega/\omega_p, \varepsilon =\nu/\omega_p $, there is such 
domain $D^+$ (its exterior is domain $D^-$), such, that if a point 
$(\Omega, \varepsilon)\in D^-$ Debay mode is absent.

Under eigen solution of initial system its general solution is obtained.
By means of boundary conditions  in an explicit form 
expressions for coefficients of discrete and continuous spectra are found.
Then in explicit form (without quadratures) absorption quantity
of energy of electric field in  slab of degenerate plasmas is found.

\end{document}